\documentclass[journal]{IEEEtran}
\usepackage[T1]{fontenc}
\usepackage{color}
\newcommand{\cy}[1]{\textcolor{red}{#1}}

\ifCLASSINFOpdf
\else
\fi
\usepackage{amsmath}

\newtheorem{proposition}{Proposition}
\newtheorem{proof}{Proof}[section]
\usepackage[justification=centering]{caption}
\usepackage{algorithmicx,algorithm}
\usepackage{algpseudocode}
\usepackage{multirow}
\usepackage{tabularx}
\usepackage{booktabs}
\usepackage{graphicx, caption, subcaption}
\usepackage{url}
\usepackage{amsmath}
\usepackage{amsfonts}
\usepackage{comment} 
\usepackage{makecell}
\usepackage{bbm}

\begin{document}
	%
	\title{A New Perspective on Stabilizing GANs training: Direct Adversarial Training}
	%
	%
	%
	
	\author{Ziqiang Li,
		Pengfei Xia,
		Rentuo Tao,
		Hongjing Niu,	
		Bin Li ~\IEEEmembership{Member,~IEEE}
		\thanks{The work is partially supported by the National Natural Science Foundation of China under grand No.U19B2044, No.61836011. (Corresponding author: Bin Li.)}
		\thanks{Ziqiang Li, Pengfei Xia, Rentuo Tao, and Hongjing Niu are with the University of Science and Technology of China, Anhui, China. (E-mail: \{iceli,xpengfei,trtmelon,sasori\}@mail.ustc.edu.cn)}
		\thanks{Bin Li (Corresponding author) is with the University of Science and Technology of China, Anhui, China. (E-mail: binli@ustc.edu.cn)}
	}
	
	%
	%

	\markboth{IEEE Transactions on Emerging Topics in Computational Intelligence}%
	{Shell \MakeLowercase{\textit{et al.}}: Bare Demo of IEEEtran.cls for IEEE Journals}
	%



	\makeatletter
	\def\ps@IEEEtitlepagestyle{%
		\def\@oddfoot{\mycopyrightnotice}%
		\def\@oddhead{\hbox{}\@IEEEheaderstyle\leftmark\hfil\thepage}\relax
		\def\@evenhead{\@IEEEheaderstyle\thepage\hfil\leftmark\hbox{}}\relax
		\def\@evenfoot{}%
	}
	\def\mycopyrightnotice{%
		\begin{minipage}{\textwidth}
			\centering \scriptsize
			Copyright~\copyright~2022 IEEE. Personal use of this material is permitted. Permission from IEEE must be obtained for all other uses, in any current or future media, including\\reprinting/republishing this material for advertising or promotional purposes, creating new collective works, for resale or redistribution to servers or lists, or reuse of any copyrighted component of this work in other works by sending a request to pubs-permissions@ieee.org.
		\end{minipage}
	}
	\makeatother
	\maketitle
	
	\begin{abstract}
		Generative Adversarial Networks (GANs) are the most popular image generation models that have achieved remarkable progress on various computer vision tasks. However, training instability is still one of the open problems for all GAN-based algorithms. Quite a number of methods have been proposed to stabilize the training of GANs, the focuses of which were respectively put on the loss functions, regularization and normalization technologies, training algorithms, and model architectures. Different from the above methods, in this paper, a new perspective on stabilizing GANs training is presented. It is found that sometimes the images produced by the generator act like adversarial examples of the discriminator during the training process, which may be part of the reason causing the unstable training of GANs. With this finding, we propose the Direct Adversarial Training (DAT) method to stabilize the training process of GANs. Furthermore, we prove that the DAT method is able to minimize the Lipschitz constant of the discriminator adaptively. The advanced performance of DAT is verified on multiple loss functions, network architectures, hyper-parameters{,} and datasets. Specifically, DAT achieves significant improvements of $11.5\%$ FID on CIFAR-100 unconditional generation based on SSGAN, $10.5\%$ FID on STL-10 unconditional generation based on SSGAN, and $13.2\%$ FID on LSUN-Bedroom unconditional generation based on SSGAN. Code will be available at \url{https://github.com/iceli1007/DAT-GAN}
	\end{abstract}
	
	\begin{IEEEkeywords}
		Adversarial training, Generative adversarial networks, Lipschitz robustness.
	\end{IEEEkeywords}

	%
	\IEEEpeerreviewmaketitle

	\section{Introduction}
	Recently, Generative Adversarial Networks (GANs) \cite{goodfellow2014generative} have been used in many generative tasks \cite{li2019af,he2021finger}, such as image inpainting \cite{yu2018generative, yeh2017semantic}, attribute editing \cite{tao2019resattr, shen2020interpreting}, and adversarial examples \cite{xiao2018generating, song2018constructing}. GANs is a two-player zero-sum game in which the discriminator measures the distance between real and generated distributions, while the generator tries to fool the discriminator by minimizing the distance. Specifically, the optimal discriminator of vanilla GAN \cite{goodfellow2014generative} estimates the $JS$ divergence, optimal discriminators of $f$-GAN \cite{nowozin2016f} and WGAN \cite{arjovsky2017wasserstein}  estimate the $f$ divergence and Wasserstein divergence, respectively. Although various types of GANs have shown impressive performance in many tasks, the training of GANs is still quite unstable and remains difficult to understand theoretically.

	There are many methods proposed to stabilize GANs training, including works of loss functions \cite{arjovsky2017wasserstein, brock2019large}, regularization and normalization technologies \cite{gulrajani2017improved, miyato2018spectral}, training algorithms \cite{heusel2018gans,yaz2018unusual}, and model architectures \cite{karras2017progressive,karras2019style,lin2022evolutionary}. Notably, Wasserstein GAN (WGAN) \cite{arjovsky2017wasserstein} is proposed to avoid the instability caused by mismatched generator and data distribution supports. The optimal discriminator of WGAN can be considered as estimating the Wasserstein distance of two distributions. To solve the Kantorovich duality problem \cite{kantorovich2006problem} of the Wasserstein distance, the discriminator should meet the 1-Lipschitz continuity, which is the first time that Lipschitz continuity has been introduced into the design of GANs. Except for 1-Lipschitz continuity in WGANs, Lipschitz continuity is also important for generalizability and distributional consistency for all GANs \cite{qi2020loss}. Qi \cite{qi2020loss} proved that, for the discriminator with Lipschitz continuity, the generated distributions converge to the real distributions in GANs. In practice, gradient penalties \cite{gulrajani2017improved} and spectral normalization \cite{miyato2018spectral} are used to implement the Lipschitz continuity. Furthermore, recent methods \cite{yang2021data,chen2019self} stabilize GANs training through representation learning of the discriminator. See \cite{li2020systematic,li2022comprehensive} for some more thorough reviews.
	
\begin{figure}
		\includegraphics[width=0.45\textwidth]{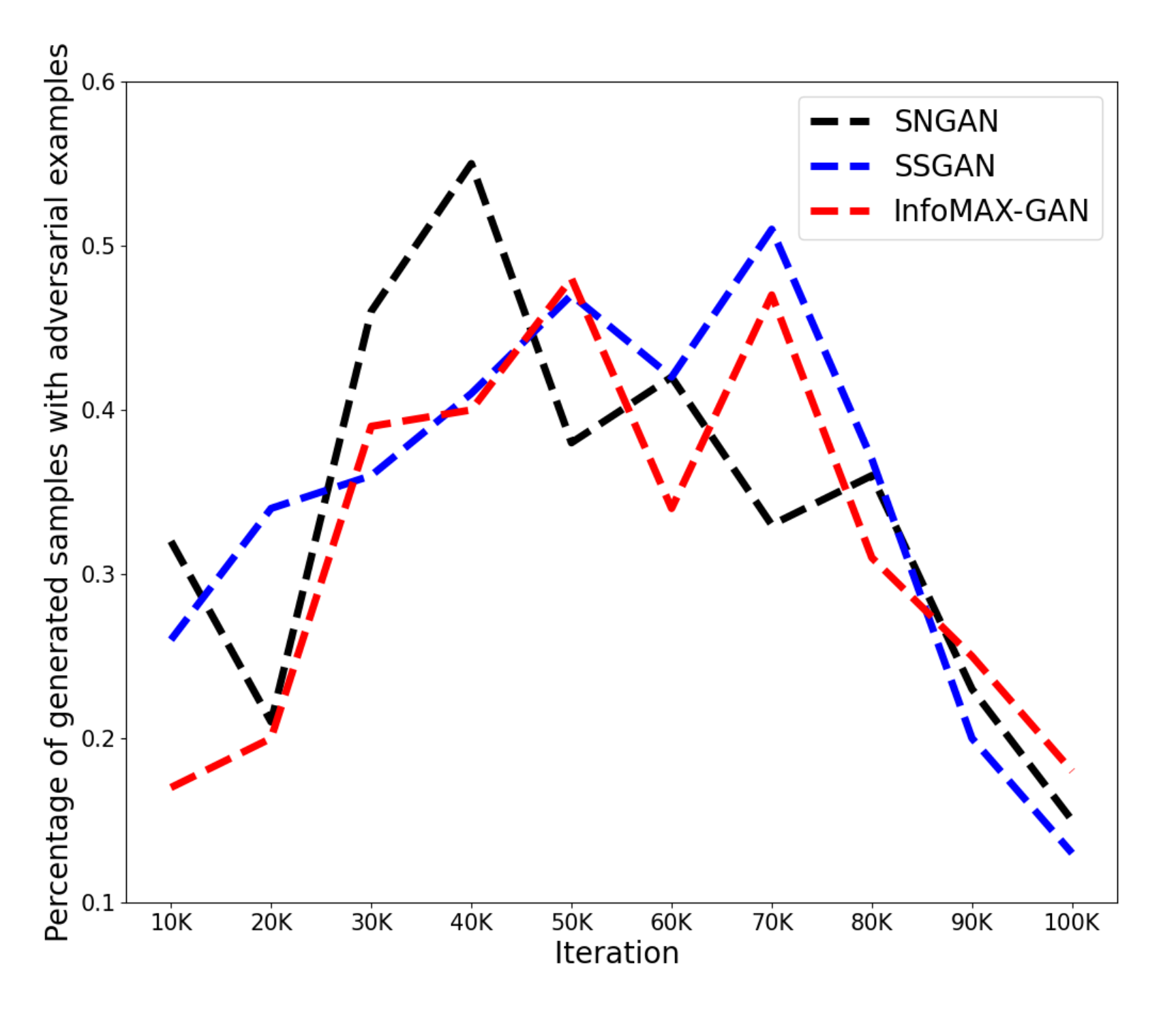}
		\centering
		\caption{Following the definition of adversarial examples, we set $\delta=0.1$ in this figure. The proportion of generated samples existing adversarial samples is sufficiently large and unstable for various GAN models, which affects the training of the GANs. All results are the average of three independent runs. }
		\label{fig:number_adversarial_examples}
	\end{figure}

	{Different from the above perspectives, we argue that the stability of GANs' training relates to the adversarial robustness of the discriminator. Non-robust discriminators are vulnerable to adversarial examples, adding noises into the gradient of the generator, which influences the stability of GANs training. Adversarial examples \cite{szegedy2013intriguing} of the discriminator are generated unintentionally by the generator during GANs training. Unrealistic adversarial perturbations in adversarial examples can incur error predictions for the discriminator. Formally, given a generated sample $x_f=G(z)$ that satisfies: $D(x_f) =F$, the adversarial examples of it are defined as $\hat x_f=G(\hat z)$ that should satisfy: $D(\hat x_f)=R \quad \text{and} \quad \|\hat x_f -x_f\|_p\leq \delta$ where $D$ and $G$ represent discriminator and generator, respectively. $F$ and $R$ are target of the discriminator's outputs for fake and real images, respectively. $\delta$ is to restrict the $\hat x_f$ to the p-norm neighbourhood of $x_f$. Although $x_f$ and its adversarial examples $\hat x_f$ are both generated by the generator and p-norm similar, the discriminator has different outputs for them. Since the model is assumed to have infinite capacity \cite{goodfellow2014generative}, without any prior, the generator has a high probability of generating adversarial examples that can mislead the discriminator. Fig. \ref{fig:number_adversarial_examples} illustrates the proportion of generated samples existing adversarial examples on the CIFAR-10 dataset.}
	
	{In this paper, a new method called Direct Adversarial Training (DAT) has been proposed to mitigate the harm of non-robust discriminator in GANs and improve the stability of training process. Furthermore, DAT can also be considered as a new regularization method that minimizes the Lipschitz constant \textbf{adaptively}. Compared to some works on loss functions \cite{arjovsky2017wasserstein, brock2019large}, training algorithms \cite{heusel2018gans,yaz2018unusual}, and model architectures \cite{karras2017progressive,karras2019style}, the proposed method is Plug-and-Play and orthogonal to the above methods. Therefore, DAT can be applied to multiple loss functions and network architectures. }

	The main contributions can be summarized as follows:
	\begin{itemize}
		\item We testify that the generator may generate adversarial examples to the discriminator during the training of GANs and analyze the the effect of it to the training process for the first time.
		\item We propose Direct Adversarial Training (DAT) for stabilizing the training process of GANs and show that the DAT can \textbf{adaptively} minimize the Lipschitz constant of the discriminator, which is different from the gradient penalty methods proposed by previous works. 
		\item Without excessive cost, our method dramatically improves the training quality and efficiency on multiple models. 
	\end{itemize}
	
	\section{Background and Related Work}
	\subsection{Generative Adversarial Networks}
	GANs is a two-player zero-sum game, where the generator $G(z)$ transforms randomly
	sampled latent distribution $z$ into the high-fidelity image distribution $p_g$ through adversarial learning. And the discriminator $D(x)$ evaluates the distance between generated distribution $p_g$ and real distribution $p_r$. {The generator and discriminator minimize and maximize the distribution distance, respectively}. This minimax game can be expressed as follows:
	\begin{equation} 
		\begin{aligned}
			\mathop{\min}\limits_{\phi}\mathop{\max}\limits_{\theta} f_{\text{GAN}}(\phi,\theta)
			=&\mathbb{E}_{x\sim p_{r}}[g_1(D_\theta(x))]\\
			+&\mathbb{E}_{z\sim p_z}[g_2(D_\theta(G_\phi(z)))],
			\label{Eq:2}
		\end{aligned}
	\end{equation}
	where $\phi$ and $\theta$ are parameters of $G$ and $D$, respectively. Specifically, vanilla GAN \cite{goodfellow2014generative} can be indicated by $g_1(t)=g_2(-t)=-\log(1+e^{-t})$, WGAN \cite{arjovsky2017wasserstein} and $f$-GAN \cite{nowozin2016f} can be demonstrated by $g_1(t)=g_2(-t)=t$ and $g_1(t)=-e^{-t},\\g_2(t)=1-t$, respectively.
	\subsection{Lipschitz Constant and WGAN}
	Lipschitz constant of the function $f: X\rightarrow Y$ is defined by:
	\begin{equation}
		||f||_L=\mathop{\sup}\limits_{x,y\in X;x\neq y}\frac{||f(x)-f(y)||}{||x-y||}.
	\end{equation}
	For a given constant $K\geq 0$ and any variables $x,y\in X$, the function $f$ satisfies the K-Lipschitz continuity when and only when $||f(x)-f(y)||\leq K||x-y||$. Furthermore, Lipschitz constant of the neural network can be approximated by spectral norm of the weight matrix \cite{li2020systematic}:
	\begin{equation}
		\|W\|_2=\mathop{\max}\limits_{x\neq 0} \frac{\|Wx\|}{\|x\|}.
	\end{equation}
	
	{The Lipschitz constant can express the Lipschitz continuity of the neural network.} The lower Lipschitz constant means that the neural network is less sensitive to input perturbation and the bounded Lipschitz constant indicates that the network has better generalization \cite{yoshida2017spectral,oberman2018lipschitz,couellan2019coupling}.
	
	For GANs, lots of works are proposed to limit the Lipschitz constant of the discriminator, such as spectral normalization \cite{miyato2018spectral} and WGAN \cite{arjovsky2017wasserstein}. Spectral normalization normalizes the spectral norm of the discriminator, which limits its Lipschitz constant to 1. {Furthermore, the Lipschitz constant in WGAN is derived from Kantorovich duality} \cite{kantorovich2006problem}, and the Wasserstein distance corresponding to the optimal transportation is represented as:
	\begin{equation}
		W(P_1,P_2)=\mathop{\sup}\limits_{||f||_L=1}\mathbb{E}_{x\sim p_{r}}f(x)-\mathbb{E}_{x\sim p_{g}}f(x),
	\end{equation}
	where $f: X\rightarrow \mathbb{R}$ is called the Kantorovich potential, which can be used as a discriminator. To make the discriminator satisfy the Lipschitz continuity, WGAN \cite{arjovsky2017wasserstein} uses the weight clipping that restricts the maximum value of each weight; WGAN-GP \cite{gulrajani2017improved} uses the gradient penalty ($\nabla_{\hat x}D_\theta(\hat x)$) with the interpolation of real samples and generated samples: $\hat{x}=tx+(1-t)y$ for $t\sim U[0,1]$ and $x\sim P_r,y\sim P_g$ being a real and generated samples; WGAN-ALP \cite{terjek2019adversarial} inspired by Virtual Adversarial Training (VAT) \cite{miyato2018virtual} restricts the 1-Lipschitz continuity at $\hat{x}=\{x,y\}$ with the direction of adversarial perturbation. Different from the above methods which restrict the 1-Lipschitz continuity \cite{gulrajani2017improved,terjek2019adversarial,kodali2017convergence,zhou2019towards}, WGAN-LP \cite{petzka2017regularization} restricts the $k$-Lipschitz continuity ($k\leq1$), which is derived from the optimal transport with regularization. Also, Qi. \cite{qi2020loss} is motivated to have lower sample complexity by directly minimizing the Lipschitz constant rather than constraining it to 1, which can be described as 0-GP \cite{zhou2019lipschitz,mescheder2018training,thanh2019improving}. {Among them, \cite{mescheder2018training} demonstrates that adding 0-GP with real ($\nabla_{ x_r}D_\theta( x_r)$) or fake images ($\nabla_{ x_f}D_\theta( x_f)$) leads to convergence of the GANs training. In summary, there are many methods \cite{li2020systematic} for restricting the Lipschitz constant, some restrict the constant to 1, some restrict it to $k$ $(k\leq 1)$, and some minimize the Lipschitz constant. See \cite{li2020systematic,li2022comprehensive} for some more thorough reviews.}
	\subsection{Adversarial Examples and Adversarial Training}
	Adversarial example is a common problem in neural networks. Given a pre-trained model $h$, adversarial examples $x'$ are defined by $x'=x+\delta$ with $h(x')\neq h(x)$ for untargeted
	attack or $h(x')=t$ for targeted attack, where $x$ is a clean image and $\delta$ is an imperceptible tiny perturbation. There are many methods to get adversarial examples, such as Fast Gradient Sign Method (FGSM) \cite{goodfellow2014explaining}, Projected Gradient Descent (PGD) \cite{madry2017towards}, and Basic Iterative Methods (BIM) \cite{kurakin2016adversarial}. FGSM uses the single gradient step to generate adversarial examples:
	\begin{equation}
		\left\{
		\begin{aligned}
			x'&=x+\epsilon\cdot \operatorname{sign}(\nabla_x\mathcal{L}(x,y))\quad\quad \operatorname{for\ untargeted},\\
			x'&=x-\epsilon\cdot \operatorname{sign}(\nabla_x\mathcal{L}(x,t))\quad\quad \operatorname{for\ targeted},
		\end{aligned}
		\right.
		\label{Eq:6}
	\end{equation}
	where  $\mathcal{L}$ is the loss function of the classification. PGD is a multi-step method that creates adversarial examples by iterative:
	\begin{equation}
		\left\{
		\begin{aligned}
			x^{k+1}&=\mathbf{clip}\big( x^k+\alpha  \cdot \operatorname{sign}(\nabla_x\mathcal{L}(x^k,y))\big)\quad \operatorname{for\ untargeted},\\
			x^{k+1}&=\mathbf{clip}\big( x^k-\alpha\cdot \operatorname{sign}(\nabla_x\mathcal{L}(x^k,t))\big)\quad \operatorname{for\ targeted},
		\end{aligned}
		\right.
	\end{equation}
	where $\mathbf{clip}$ is the clip function, $\alpha$ is the step size of gradient, $x^0=x$, $x'=x^K$, and $K$ is the number of iterations. Besides, some works \cite{hu2017generating,samangouei2018defense} use GANs to generate adversarial examples.
	
	Adversarial training is a good and simple method to avoid adversarial examples, which improves the robustness of the neural networks by introducing adversarial examples into training:
	\begin{equation}
		\mathop{\min}_\theta \mathbb{E}_{x,y\sim D}[\mathop{\max}_{||\delta||_p\leq\epsilon}\mathcal{L}_\theta(x+\delta,y)],
	\end{equation}
	where $x,y\sim D$ are sampled from the joint distribution of data (image, label), $\theta$ is the parameters of the network. This min-max problem is similar to the GANs. {The main difference between them is that the independent variable of the maximization problem in the adversarial training is image samples $x$, rather than the discriminator parameters.}
	\subsection{The Adversarial Robustness with GANs Training}
	Recently, some works {have begun} to analyze the relationship between GANs training and adversarial robustness. RobGAN \cite{liu2019rob} is the first work to introduce adversarial training in GANs training, which adds the adversarial training for the classifier in cGANs. This method does not directly analyze adversarial examples of the discriminator. Furthermore, {Zhou \textit{et al.}} \cite{zhou2018don} analyze the non-robust characteristics of the discriminator for the first time. They propose the consistent regularization between adversarial images and clean images during the training of GANs, which will let the discriminator not be fooled by adversarial examples. As we all know, adversarial training is a remarkable method to improve the robustness of networks. Concurrent with our work, several methods \cite{zhang2020robust,liu2020adversarial} independently propose adversarial training for the training of GANs. Compared with concurrent works mentioned above, our paper analyzes the possibility of the generator to generate adversarial examples and the relationship between Direct Adversarial Training and adaptive Lipschitz minimum. Moreover, our work contains richer experiment results and better performance.

	\section{Proposed Approach}
	The pipeline of DAT is applying adversarial examples to train GANs, which is the same as other adversarial training methods on classifier. However, unlike previous adversarial training for classifiers, the proposed DAT method targets to a distribution metric function: discriminator. The unintentional attack for discriminator can be regarded as a targeted attack that minimizes the distance (under the discriminator) between the adversarial examples of the real images and the generated images, the same is true for adversarial examples of the fake images.
	
	In the following contents, we first introduce the DAT method, from which an adversarial perturbation strategy according to various distribution metrics have been proposed. Moreover, we also analyze the correlation between the proposed method and the gradient penalty.
	\begin{algorithm*}[t]
		\caption{Direct Adversarial Training} 
		\label{algorithm:1}
		\hspace*{0.08in} {\bf Input:}
		The batch size m, the real image distribution $P_r(x)$, the random noize $z\sim N(0,1)$, the maximum number of training steps K, the number of
		steps to the discriminator N, the loss function $g_1$ and $g_2$.           \\
		\hspace*{0.08in} {\bf Output:}
		a fine-tuned generator G and discriminator D
		\begin{algorithmic}[1]
			\For{k=1, 2, $\cdots$, K} 
			\For{n=1,2, $\cdots$, N}
			\State Sampled m real samples $x_r=\{x^{(1)}_r, x^{(2)}_r, \cdots, x^{(m)}_r\}$ from the real distribution $p_r$.
			\State Sampled m latent noise $z=\{z^{(1)}, z^{(2)}, \cdots, z^{(m)}\}$ from the low-dimension latent distribution $p_z$.
			\State $x_f=\{x^{(1)}_f, x^{(2)}_f, \cdots, x^{(m)}_f\}=G_{\phi}\big(\{z^{(1)}, z^{(2)}, \cdots, z^{(m)}\}\big)$
			\State $\delta^{(i)}_f=-\epsilon\nabla_{x^{(i)}_f}\big(\left|g_1(D(x^{(i)}_f))-\overline{ g_1(D(x_r))}\right|\big)$ for $i\in{1,2,\cdots,m}$
			\State $\delta^{(i)}_r=-\epsilon\nabla_{x^{(i)}_r}\big(\left|g_2(D(x^{(i)}_r))-\overline{ g_2(D(x_f))}\right|\big)$ for $i\in{1,2,\cdots,m}$
			\State $\hat x^{(1)}_{f}, \hat x^{(2)}_{f}, \cdots, \hat x^{(m)}_{f}=\{x^{(1)}_f+\delta^{(1)}_f, x^{(2)}_f+\delta^{(2)}_f, \cdots, x^{(m)}_f+\delta^{(m)}_f\}$
			\State $\hat x^{(1)}_{r}, \hat x^{(2)}_{r}, \cdots, \hat x^{(m)}_{r}=\{x^{(1)}_r+\delta^{(1)}_r, x^{(2)}_r+\delta^{(2)}_r, \cdots, x^{(m)}_r+\delta^{(m)}_r\}$
			\State Update the discriminator by ascending its stochastic gradient:
			$$
			\nabla_{\theta}\biggl\{\frac{1}{m}\sum_{i=1}^{m}\big[g_1(D_{\theta}(\hat x^{(i)}_{r}))+g_2(D_{\theta}(\hat x^{(i)}_{f}))\big]\biggr\}
			$$
			\EndFor
			\State Draw m latent noise $\{z^{(1)}, z^{(2)}, \cdots, z^{(m)}\}$.
			\State Update the generator by descending its stochastic gradient:
			$$
			\nabla_{\phi}\biggl\{\frac{-1}{m}\sum_{i=1}^{m}\big[g_1(D_{\theta}(x^{(i)}_r))+g_2(D_{\theta}(G_{\phi}(z^{(i)})))\big]\biggl\}
			$$
			\EndFor
			
			\State \Return 
		\end{algorithmic}
	\end{algorithm*}
	
	\subsection{Direct Adversarial Training}
\begin{figure}
		\includegraphics[width=0.45\textwidth]{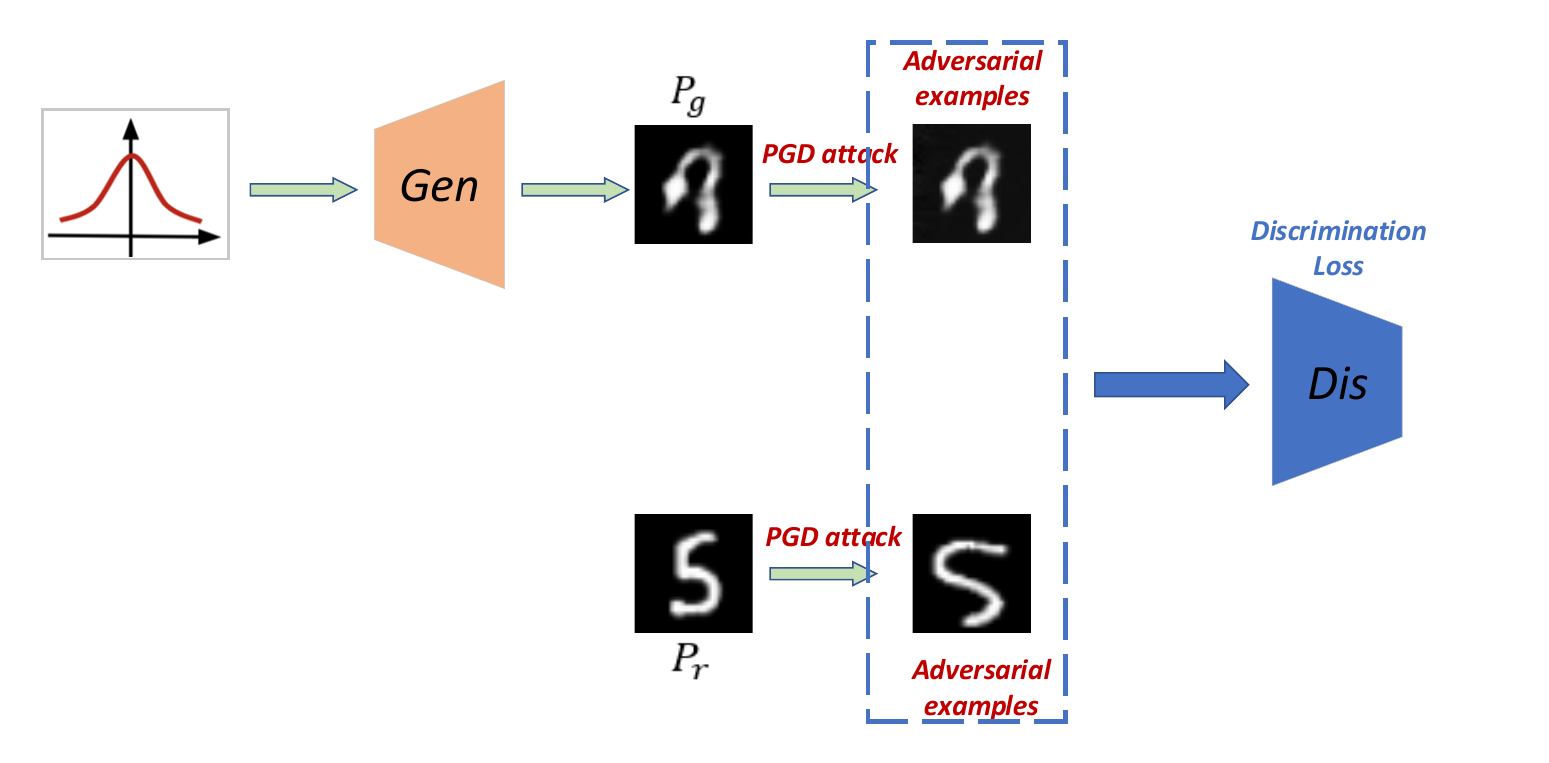}
		\centering
		\caption{Introductions of the proposed DAT that is similar to the training of GANs. The difference is that we add the one-step PGD attack for real and generated images, we hope that the discriminator not only can identify real or fake, but also be robust to adversarial examples.}
		\label{fig:framework}
	\end{figure}

	Motivated by adversarial examples leading to unstable training of GANs, we propose an adversarial training method DAT for GANs in Fig.\ref{fig:framework}. According to the loss function of GANs in Eq (\ref{Eq:2}), the loss of GANs with DAT can be formed as: 
	
	\begin{equation}
		\begin{aligned} 
			\mathop{\min}\limits_{\phi}\mathop{\max}\limits_{\theta} f_{\text{DAT}}(\phi,\theta)= 
			\mathbb{E}_{x_r\sim p_{r}}[g_1(D_\theta(\hat x_r))]\\+\mathbb{E}_{x_f\sim p_f}[g_2(D_\theta(\hat x_f))],
		\end{aligned} 
	\end{equation}
	where $\hat x_r$ and $\hat x_f$ are adversarial examples of real and fake images, respectively. They are defined as $\hat x_r=x_r+\delta(x_r)$, $\hat x_f=x_f+\delta(x_f)$, and $x_f=G_\phi(z)$, where $\delta(x_r)$ and $\delta(x_f)$ are adversarial perturbation of $x_r$ and $x_f$, respectively. {In this part, we use some image perturbations to obtain adversarial examples, which is an approximation to the adversarial examples defined in Introduction. Adversarial examples defined in the Introduction can be considered generated by latent perturbations, which need too much cost to obtain in training.} According to the above formulas, the complete algorithm is demonstrated in \textbf{Algorithm 1}.
	
	\subsection{Adversarial Perturbation of the Discriminator}
	
	Most of the adversarial training is about classifiers, {and} the goal of the classifier is fixed. For untargeted attacks, the direction of adversarial perturbation is the solution of $\mathop{\max}\mathcal{L}(x,y)$. For targeted attacks, the direction of adversarial perturbation is the solution of $\mathop{\min}\mathcal{L}(x,t)$, where $\mathcal{L}$ is the loss function of the classifiers{,} and $t$ is target label of the attack. But for discriminator, goal of the output is dynamic. {For instance, the outputs of the real images and generated images are both $0.5$ for the optimal discriminator in vanilla GAN, but which is not true at the beginning of training.} So the target label of adversarial attack changes dynamically with training. Based on this, we propose a new adversarial perturbation for the discriminator, in which the direction of adversarial perturbation for real images is the solution of $\mathop{\arg  \min}\left|g_1(D_\theta(x_r+\delta(x_r)))-\overline{ g_1({D}_\theta(x_f))}\right|$. Similarly, the direction of adversarial perturbation for generated images is the solution of $\mathop{\arg  \min}\left|g_2(D_\theta(x_f+\delta(x_f)))-\overline{ g_2( D_\theta(x_r))}\right|$. Taking {the} above together, the adversarial perturbation for the discriminator can be defined as:
	\begin{equation}
		\begin{aligned}
			\delta(x_{r})=\mathop{\arg  \min}\left|g_1(D_\theta(x_r+\delta(x_r)))-\overline{ g_1({D}_\theta(x_f))}\right|,\\
			\delta(x_{f})=\mathop{\arg  \min}\left|g_2(D_\theta(x_f+\delta(x_f)))-\overline{ g_2( D_\theta(x_r))}\right|,
			\label{Eq:9}
		\end{aligned}
	\end{equation}
	where $x_{r}$ and $x_{f}$ are real image and generated image, respectively. $\delta(x_r)$ and $\delta(x_f)$ are adversarial perturbation of the real image $x_r$ and generated image $x_f$, respectively. $\overline{ g_1(D_\theta(x_{r}))}$ and $\overline{ g_2( D_\theta(x_{f}))}$ are the average of the discriminator output in batches for the real and generated images, which can be considered as target label of the generated and real images attack, respectively. {$\left|\cdot\right|$} indicates the calculation of absolute value. For above optimization problems, we use the one-step PGD attack to achieve:
	\begin{equation}
		\begin{aligned}
			\delta(x_r)=-\epsilon\nabla_{x_r}\big(\left|g_1(D_\theta(x_r))-\overline{ g_1(D_\theta(x_f))}\right|\big),\\
			\delta(x_f)=-\epsilon\nabla_{x_f}\big(\left|g_2(D_\theta(x_f))-\overline{ g_2(D_\theta(x_r))}\right|\big),
			\label{Eq:17}
		\end{aligned}
	\end{equation}
	where $\epsilon$ is perturbation level\footnote{Generally, we set $\epsilon=1$, while more discussion of the $\epsilon$ will be given in section 6.1} and the goal of the adversarial perturbation changes with the training of the discriminator.

	\subsection{Direct Adversarial Training and Lipschitz Continuity}
	{Minimizing the Lipschitz constant through gradient penalty could stabilize the training of the discriminator, thus ensuring the convergence of the GANs \cite{mescheder2018training}. However, recent studies \cite{wu2021gradient} demonstrate that the limitation restricts the capacity of the discriminator and thus deteriorates the performance of the generative model.} In this part, we analyze the relationship between DAT and gradient penalty. The result demonstrates that our method can minimize the Lipschitz constant of the discriminator adaptively. 
	The instability of GANs training is mainly caused by the discriminator, and our adversarial training is only for the discriminator, so we do not consider the generator at present. For adversarial perturbation of real images $\delta(x_r)=-\epsilon\cdot\nabla_{x_r}\big(\left|g_1(D(x_r))-\overline{ g_1( D(x_f))}\right|\big)$, the loss function of the discriminator can be written as:
	\begin{equation}
		\begin{aligned}
			&\mathop{\max}_{\theta}\mathbb{E}_{x\sim p_r}\bigg[g_1\big(D_\theta(x+\delta{(x)})\big)\bigg]\\
			\approx&\mathop{\max}_{\theta}\mathbb{E}_{x\sim p_r}\bigg[ g_1\big(D_\theta(x)\big)+\nabla_{x}g_1\big(D_\theta(x)\big)\cdot \delta(x)\bigg].
			\label{Eq:11}
		\end{aligned}
	\end{equation}
	Considering the interior of expectation in the Eq (\ref{Eq:11}):
	\begin{equation}
		\begin{aligned}
			& g_1\big(D_\theta(x_r)\big)+\nabla_{x_r}g_1\big(D_\theta(x_r)\big)\cdot \delta(x_r)\\
			=& g_1\big(D_\theta(x_r)\big)-\epsilon\nabla_{x_r}g_1\big(D_\theta(x_r)\big)\\
			&\cdot \nabla_{x_r}\left|g_1(D_\theta(x_r))-\overline{ g_1(D_\theta(x_f))}\right|\\
			=&g_1\big(D_\theta(x_r)\big)-\epsilon g_1'\nabla_{x_r}D_\theta(x_r)\\
			&\cdot \nabla_{x_r}\left|g_1(D_\theta(x_r))-\overline{ g_1( D_\theta(x_f))}\right|.
			\label{Eq:12} 
		\end{aligned}
	\end{equation}
	{Because $x_f$ is sampled from fake images, which is independent of $x_r$, when $g_1(D_\theta(x_r)) - \overline{ g_1( D_\theta(x_f))}\geq 0$, it is true in most cases, the Eq (\ref{Eq:12}) is {$g_1\big(D_\theta(x_r)\big)-\epsilon {g_1'}^2\|\nabla_{x_r}D_\theta(x_r)\|_2^2$}. In this case, the loss function of the discriminator to real imgaes is represented as $\mathop{\max}\limits_{\theta}\bigg[ g_1\big(D_\theta(x_r)\big)-\epsilon {g_1'}^2\|\nabla_{x_r}D_\theta(x_r)\|_2^2 \bigg]$, whch is equivalent to adding gradient penalty (0-GP) to loss function. 0-GP can be used to limit the Lipschitz constant and stabilize the training of GANs. However, when $g_1(D_\theta(x_r))-\overline{ g_1( D_\theta(x_f))}< 0$, which means that the discriminator is incorrect to the $x_r$. In this case, the loss function of the discriminator to real imgae $x_r$ is represented as $\mathop{\max}\limits_{\theta}\bigg[g_1\big(D_\theta(x_r)\big)+\epsilon {g_1'}^2\|\nabla_{x_r}D_\theta(x_r)\|_2^2$. In this case, $\mathop{\max}\limits_{\theta} \|\nabla_{x_r}D_\theta(x_r)\|_2^2 $ means that we hope the discriminator will have a large change for perturbation (gradient reward), so as to jump out of the situation of the wrong discrimination.}
	
	{In summary, the proposed DAT can adaptively choose strategies (gradient penalty or gradient reward) for different samples. Therefore, the loss function of the discriminator to real images in DAT can be summarized as:}
	
	\begin{equation}
		\left\{
		\begin{aligned}
			\mathop{\max}_{\theta}&\bigg[ g_1\big(D_\theta(x_r)\big)-\epsilon {g_1'}^2\|\nabla_{x_r}D_\theta(x_r)\|_2^2 \bigg] \\
			&\text{for}\quad g_1(D_\theta(x_r)) - \overline{ g_1( D_\theta(x_f))}\geq 0 \quad\text{Gradient Penalty}\\
			\mathop{\max}\limits_{\theta}&\bigg[g_1\big(D_\theta(x_r)\big)+\epsilon {g_1'}^2\|\nabla_{x_r}D_\theta(x_r)\|_2^2\bigg]\\
			&\text{for}\quad g_1(D_\theta(x_r)) - \overline{ g_1( D_\theta(x_f))}< 0 \quad\text{Gradient Reward}
		\end{aligned}
		\right.
	\end{equation}
	
	For adversarial perturbation of generated images $\delta(x_f)=-\epsilon\nabla_{x_f}\big(\left|g_2(D(x_f))-\overline{ g_2( D(x_r))}\right|\big)$, where $x_f=G_\phi(z)$. Update of the discriminator can be written as:
	\begin{equation}
		\begin{aligned}
			&\mathop{\max}_{\theta}\mathbb{E}_{x_f\sim p_f}\bigg[g_2\big(D_\theta(x_f+\delta{(x_f)})\big)\bigg]\\
			\approx&\mathop{\max}_{\theta}\mathbb{E}_{x_f\sim p_f}\bigg[ g_2\big(D_\theta(x_f)\big)+\nabla_{x}g_2\big(D_\theta(x_f)\big)\cdot \delta(x_f)\bigg].
			\label{Eq:14}
		\end{aligned}
	\end{equation}
	\begin{figure*}
		\begin{subfigure}[b]{0.15\textwidth}
			\centering
			\includegraphics[width=\textwidth]{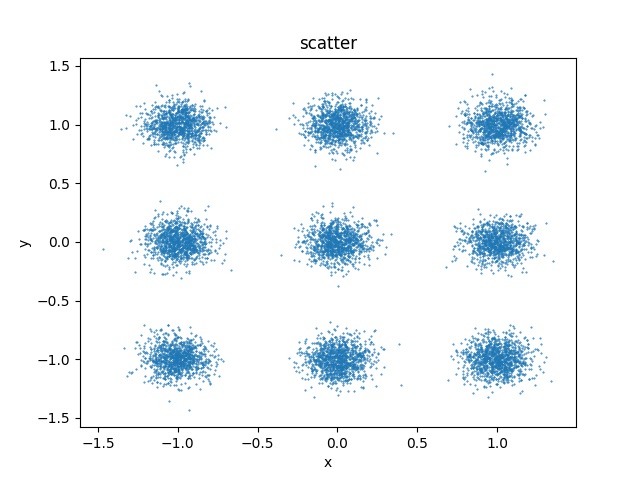}
			\caption{real samples}
			
		\end{subfigure}\hspace{-3mm}
		\begin{subfigure}[b]{0.15\textwidth}
			\centering
			\includegraphics[width=\textwidth]{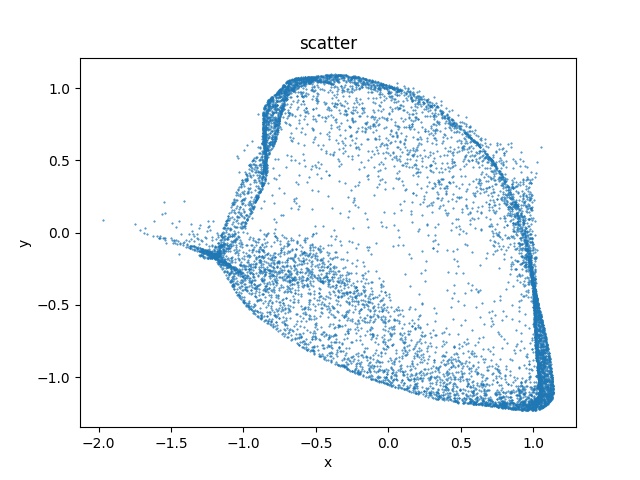}
			\caption{GAN-2k}
			
		\end{subfigure}\hspace{-3mm}
		\begin{subfigure}[b]{0.15\textwidth}
			\centering
			\includegraphics[width=\textwidth]{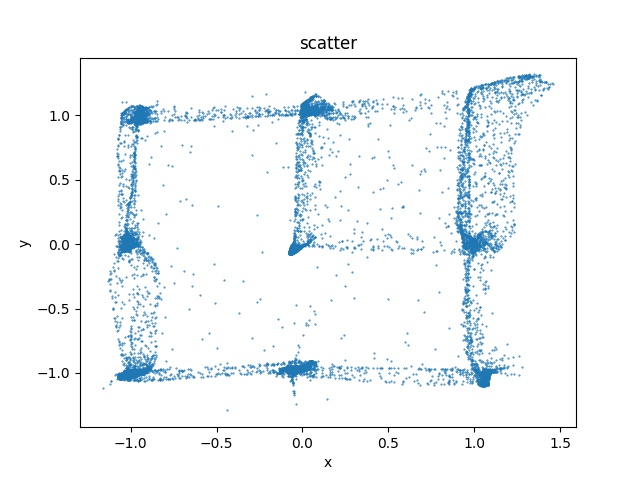}
			\caption{GAN-3k}
			
		\end{subfigure}\hspace{-3mm}
		\begin{subfigure}[b]{0.15\textwidth}
			\centering
			\includegraphics[width=\textwidth]{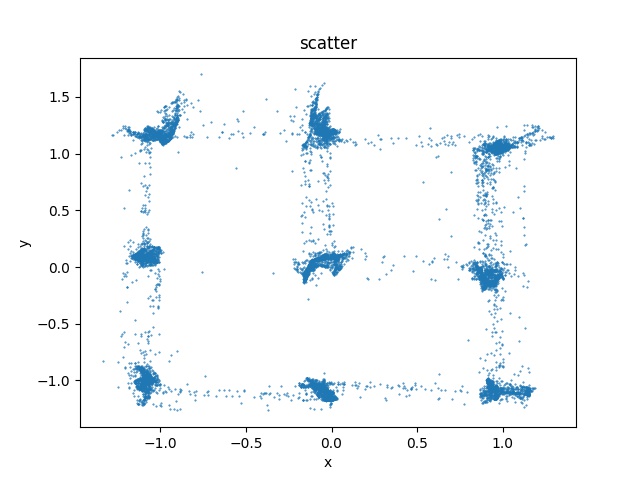}
			\caption{GAN-10k}
			
		\end{subfigure}\hspace{-3mm}
		\begin{subfigure}[b]{0.15\textwidth}
			\centering
			\includegraphics[width=\textwidth]{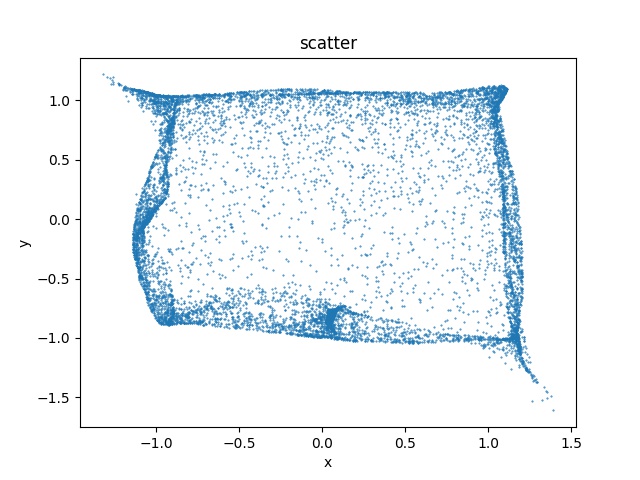}
			\caption{GAN-DAT-2k}
			
		\end{subfigure}\hspace{-3mm}
		\begin{subfigure}[b]{0.15\textwidth}
			\centering
			\includegraphics[width=\textwidth]{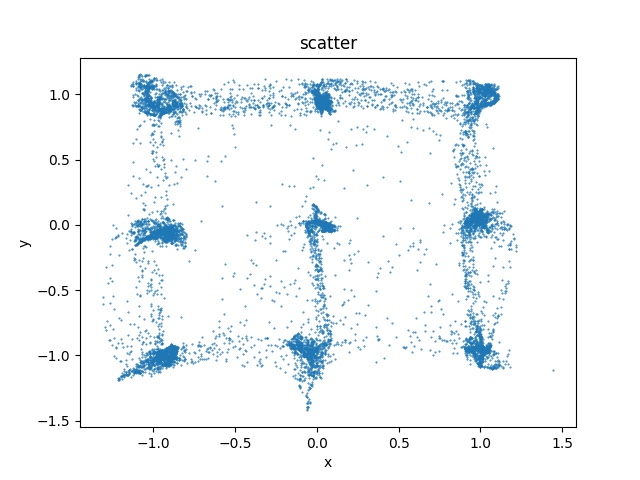}
			\caption{GAN-DAT-3k}
			
		\end{subfigure}\hspace{-3mm} 
		\begin{subfigure}[b]{0.15\textwidth}
			\centering
			\includegraphics[width=\textwidth]{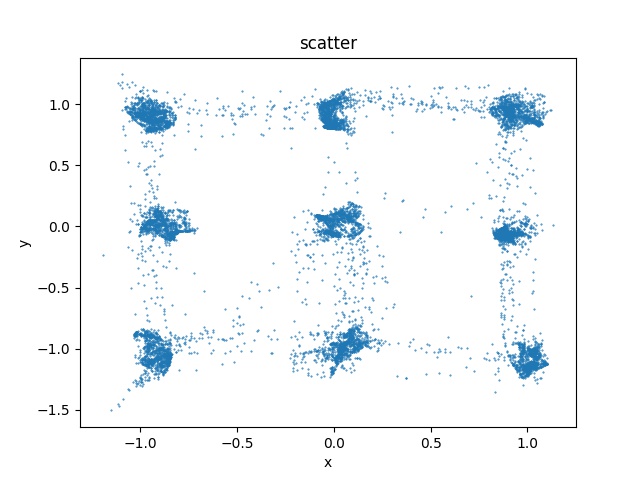}
			\caption{GAN-DAT-10k}
			
		\end{subfigure}\hspace{-3mm} \\
		\begin{subfigure}[b]{0.15\textwidth}
			\centering
			\includegraphics[width=\textwidth]{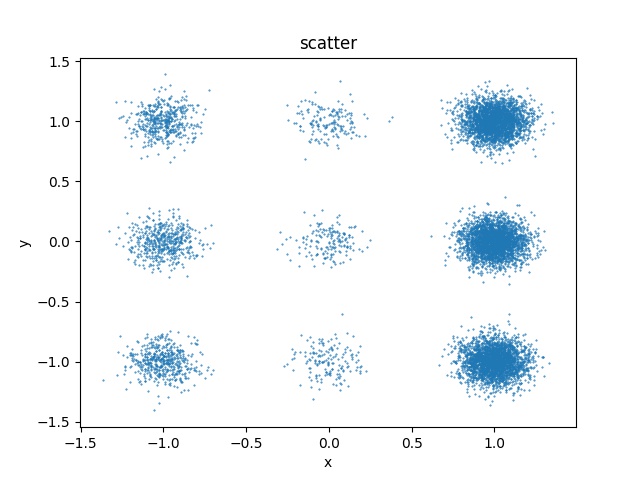}
			\caption{real samples}
			
		\end{subfigure}\hspace{-3mm}
		\begin{subfigure}[b]{0.15\textwidth}
			\centering
			\includegraphics[width=\textwidth]{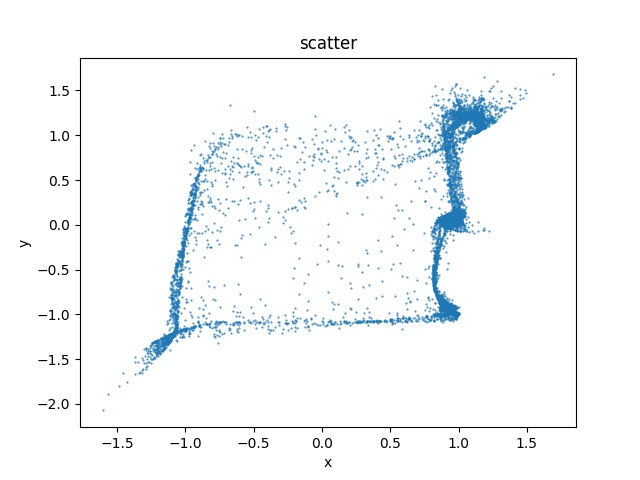}
			\caption{GAN-2k}
			
		\end{subfigure}\hspace{-3mm}
		\begin{subfigure}[b]{0.15\textwidth}
			\centering
			\includegraphics[width=\textwidth]{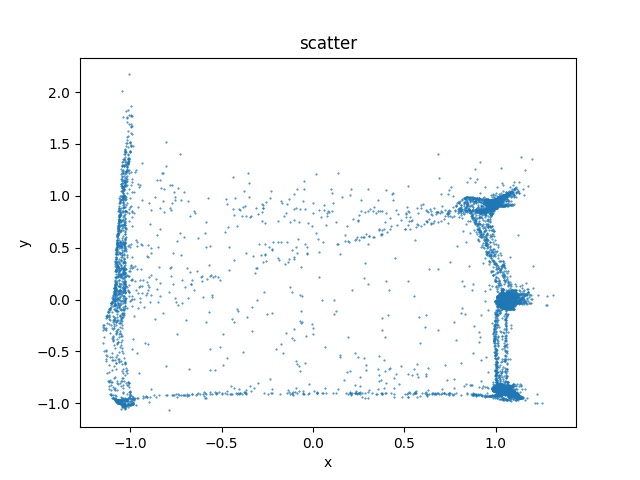}
			\caption{GAN-3k}
			
		\end{subfigure}\hspace{-3mm}
		\begin{subfigure}[b]{0.15\textwidth}
			\centering
			\includegraphics[width=\textwidth]{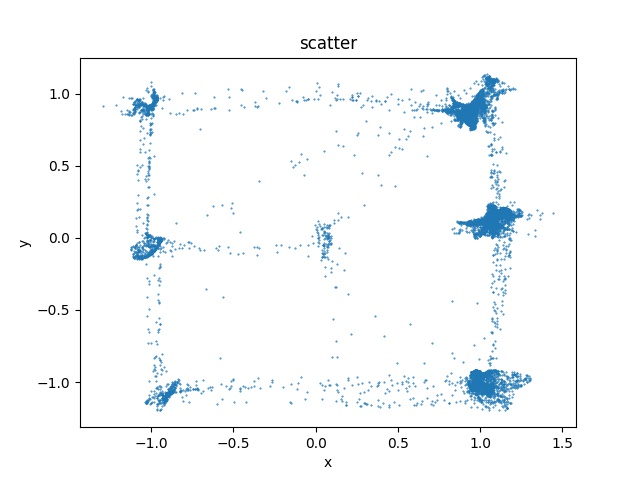}
			\caption{GAN-10k}
			
		\end{subfigure}\hspace{-3mm}
		\begin{subfigure}[b]{0.15\textwidth}
			\centering
			\includegraphics[width=\textwidth]{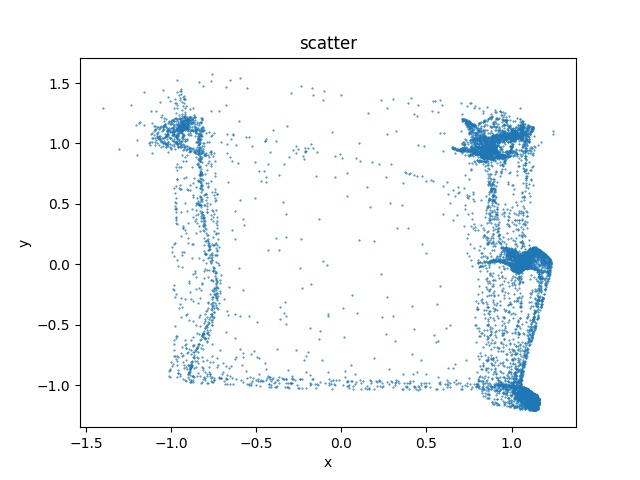}
			\caption{GAN-DAT-2k}
			
		\end{subfigure}\hspace{-3mm}
		\begin{subfigure}[b]{0.15\textwidth}
			\centering
			\includegraphics[width=\textwidth]{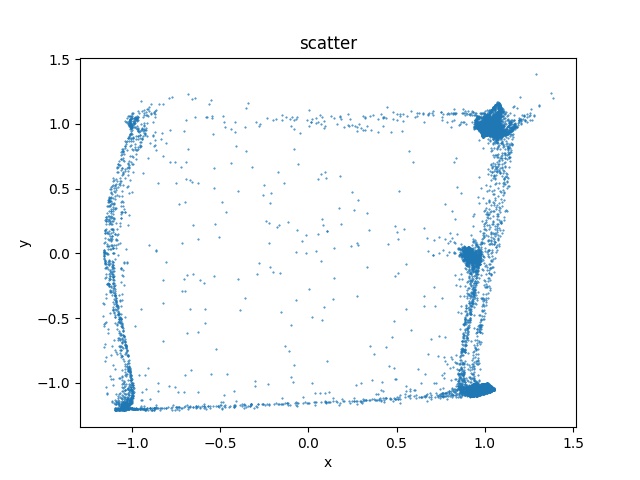}
			\caption{GAN-DAT-3k}
			
		\end{subfigure}\hspace{-3mm} 
		\begin{subfigure}[b]{0.15\textwidth}
			\centering
			\includegraphics[width=\textwidth]{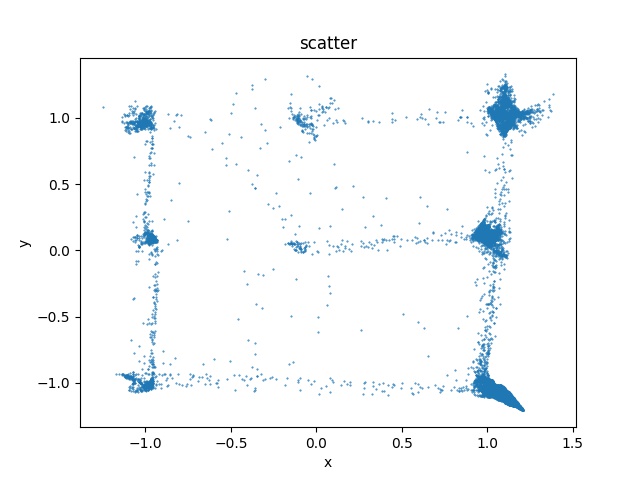}
			\caption{GAN-DAT-10k}
			
		\end{subfigure}
		\caption{Qualitative results of nine 2D-Gaussian synthetic data. The first line is a \textbf{balanced} Gaussian mixture distribution, and the second line is an \textbf{imbalanced} Gaussian mixture distribution.
			\textbf{First line:} (a): the real samples from a balanced mixture of nine Gaussian distributions, where variance is 0.1 and means are \{-1, 0, 1\};
			(b), (c) and (d): the results of iterating 2k times, 3k times, and 10k times in DCGAN; (e), (f) and (g): the results of iterating 2k times, 3k times, and 10k times in DCGAN with DAT. \textbf{Second line:} (h): the real samples from an imbalanced mixture of nine Gaussians. The probability of mixing Gaussian from left to right is 0.15, 0.05, 0.8. The other parts are the same as the first line.}
		\label{figure:real_samples}
	\end{figure*}
	We can get the loss function similar to the real images:
	\begin{equation} 
		\begin{aligned}
			g_2\big(D_\theta(x_f)\big)-&\epsilon g_2'\nabla_{x_f}D_\theta(x_f)\\
			&\cdot \nabla_{x_f}\left|g_2(D_\theta(x_f))-\overline{ g_2( D_\theta(x_r))}\right|.
			\label{Eq:15}
		\end{aligned}
	\end{equation}
	When $g_2(D_\theta(x_f))\geq\overline{ g_2( D_\theta(x_r))}$, the Eq (\ref{Eq:15}) is $g_2\big(D_\theta(x_f)\big)-\epsilon {g_2'}^2\|\nabla_{x_f}D_\theta(x_f)\|_2^2$. For generated images $x_f$, usually, $g_2(D_\theta(x_f))$ is greater than $g_2(D_\theta(x_r))$, in this case the adversarial training is equivalent to 0-GP; and $g_2(D_\theta(x_f))<\overline{ g_2( D_\theta(x_r))}$ 
	indicates that the discriminator has an error and may get the local saddle point, so we maximize the gradient of the discriminator $\|\nabla_{x_f}D_\theta(x_f)\|_2^2$, which can make the discriminator jump out of the error point as soon as possible.
	
	From the above analyses, it can be seen that our DAT can adaptively minimize the Lipschitz continuity, which is equivalent to 0-GP when the discriminator performance is better, and relax the limit on the Lipschitz constant when the discriminator performance is poor.
	
	\section{Experiments}
	In this section, we implement the proposed DAT method for DCGAN, Spectral Normalization GAN (SNGAN) \cite{miyato2018spectral}, Self-Supervised GAN (SSGAN) \cite{chen2019self}, and Information Maximum GAN (InfoMAXGAN) \cite{lee2020infomax}. Furthermore, we adopt three metrics to evaluate the performance of the method: Fréchet Inception Distance (FID) \cite{heusel2017gans}, Kernel Inception Distance (KID) \cite{binkowski2018demystifying}, and Inception Score (IS) \cite{salimans2016improved}. Among them, IS only uses generated images to measure the performance of the GANs, where FID and KID evaluate the performance of GANs by measuring the distance between real and generated images. So FID and KID are calculated on both the train and test datasets. {More descriptions on FID, IS, and KID will be shown in section A of the Appendix.}
	
	We validate the method on the following datasets: 
	\begin{itemize}
		\item {This paper creates the 2D synthetic dataset that contains two tiny sets}: balanced mixture and imbalanced mixture. Both sets are sampled from a 2D mixture distribution of nine Gaussian distribution, where the variance of the Gaussian distribution is 0.1, covariance is 0, and means are \{-1, 0, 1\}. The probability of each Gaussian distribution in the balanced mixture is equal, while the probabilities of Gaussian distribution in the imbalanced mixture are 0.15, 0.05, 0.8 from left to right. The visualizations are illustrated in (a) and (h) of Figure \ref{figure:real_samples}, respectively.
		\item The CIFAR-10 datasets \cite{krizhevsky2009learning} contains 60K images with the resolution of 32 $\times$ 32 pixels, including 50K images in the train set and 10K images in the test set{,} and all images are divided equally into 10 classes. Thus train dataset is used to train GANs' models and calculate the Train FID and Train KID. Naturally, Test FID and Test KID are calculated on the test dataset. 
		\item The CIFAR-100 dataset \cite{krizhevsky2009learning} is similar to the CIFAR-10 dataset that includes 50K train images and 10K test images, but all images are divided equally into 100 classes. Thus train dataset is used to train GANs' models and calculate the Train FID and Train KID. Naturally, Test FID and Test KID are calculated on the test dataset. 
		\item The STL-10 dataset \cite{coates2011analysis} contains 100K unlabeled images and 8K test images with a spatial resolution of 96 $\times$ 96 resolution. {All images are resized to 48 $\times$ 48 pixels saving computing resources}. Furthermore, GAN models are trained on the whole unlabeled dataset{,} and Train FID and Train KID are computed on the unlabeled dataset with random 50K images. Naturally, Test FID and Test KID are calculated on the test dataset. 
		\item The Tiny ImageNet dataset \cite{wu2017tiny} includes 200 classes with 100K images for training and 10K images for testing. All images are of size 64 $\times$ 64. Similarly, we train GAN models on the whole train dataset and compute Train FID and Train KID on the training dataset with 50K random images. Naturally, Test FID and Test KID are calculated on the test dataset. 
		\item The LSUN Bedroom dataset \cite{yu2015lsun} has approximately 3M images. We took out 10K of them as a test set which is used to calculate the Test FID and Test KID and train the models on the rest images. Naturally, we compute Train FID and Train KID on the training dataset with 50K random images.
	\end{itemize}
	
	\begin{table*}
		\caption{Some parameters on SOTA methods\label{Tab:parameters}}
		\centering
		\begin{tabular}{ccccccccc}
			\toprule
			Dataset&Resolution&	Batch Size&	Learning Rate&	$\beta_1$&$\beta_2$&	Decay Policy&	$n_{dis}$&	$n_{iter}$\\
			\midrule
			CIFAR-10&32 x 32&	64&	2e-4&	0.0&	0.9&	Linear&	5&	100K\\
			CIFAR-100&32 x 32&	64&	2e-4&	0.0&	0.9&	Linear&	5&	100K\\
			STL-10&48 x 48&	64&	2e-4&	0.0&	0.9&	Linear&	5&	100K\\
			Tiny-ImageNet&64 x 64&	64&	2e-4&	0.0&	0.9&	None&	5&	100K\\
			LSUN-Bedroom&64 x 64&	64&	2e-4&	0.0&	0.9&	Linear&	5&	100K\\
			\bottomrule
		\end{tabular}
	\end{table*}
	\subsection{Experiments on DCGAN}
	In this section, we use DCGAN to experiment on two datasets (a 2D synthetic dataset and the CIFAR-10 dataset) showing the advancement of our method. The first experiment demonstrates that the proposed DAT can accelerate the training of GANs and avoid mode collapse{,} and the second experiment indicates that the proposed DAT is robust to hyper-parameters.
	
	We use a four-layer and fully-connected MLP with 64 hidden units per layer to model the generator and discriminator on the 2D synthetic dataset. Fig.\ref{figure:real_samples} illustrates the qualitative results of different iterations. The first line is visual results on a \textbf{balanced} mixture of nine Gaussians. DAT can speed up the generation. When iterating 2k times, the generated distribution with adversarial training is closer to the true distribution. Even after training 10k times, the GANs without adversarial training cannot fit the real distribution well.
	Also, we evaluate the method on an \textbf{imbalanced} mixture of nine Gaussians in the second line. The results illustrate that standard DCGAN will lose some small probability distribution, resulting in mode collapse, and this phenomenon will be significantly improved after applying DAT.
	
\begin{figure}
		\includegraphics[width=0.45\textwidth]{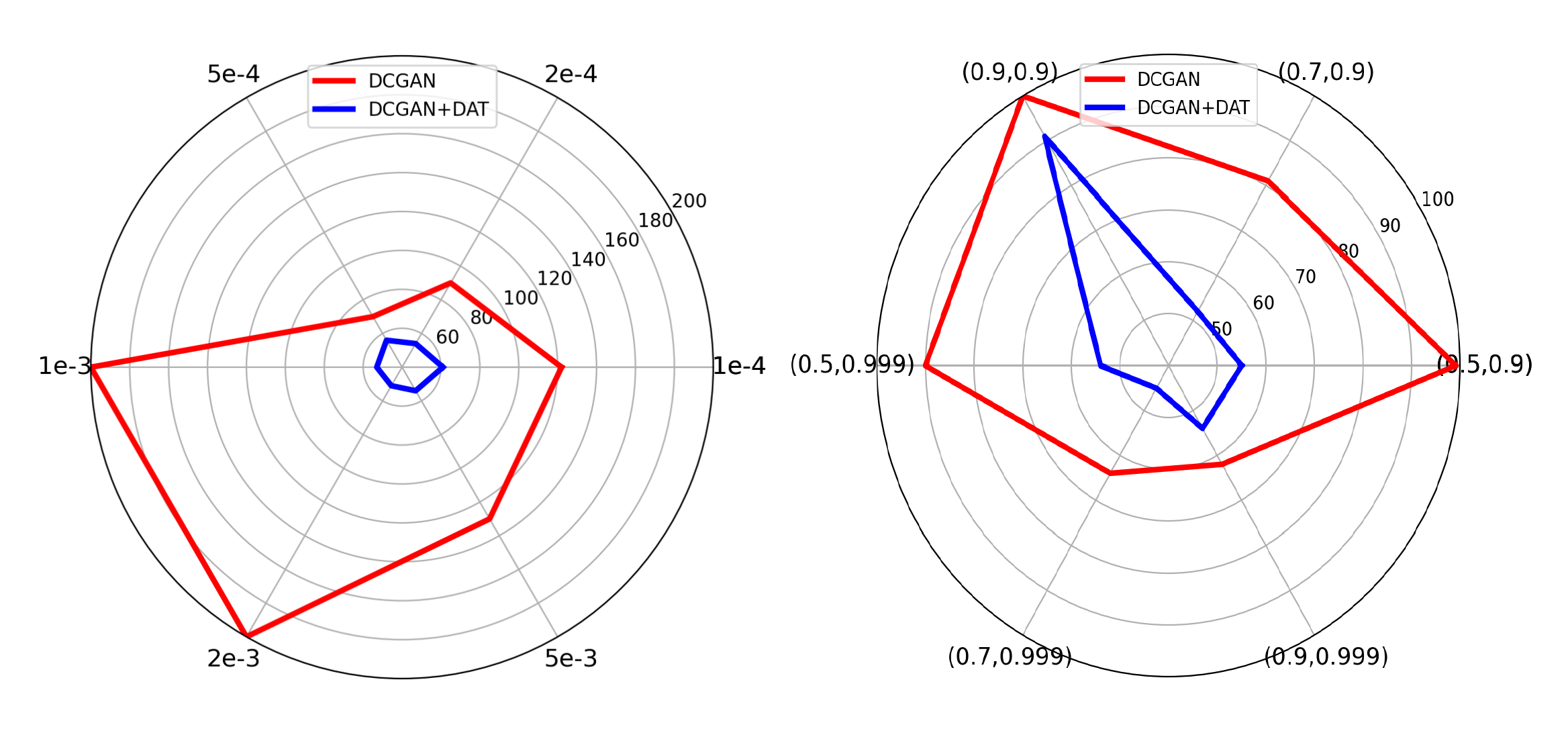}
		\centering
		\caption{Training FID of the DCGAN with different hyper-parameters on the CIFAR-10 dataset. Smaller distance to the origin indicates smaller FID/better performance. \textbf{Left:} the experiment on different Learning Rate (LR) of the Adam optimizer, where the other hyper-parameters are $\beta_1$=0.5, $\beta_2$=0.999, and $n_{dis}$=1. \textbf{Right:} the experiment on different ($\beta_1$, $\beta_2$) of the Adam optimizer, where the other hyper-parameters are LR=0.0002 and $n_{dis}$=1. Compared to the vanilla training DCGAN ({\rule[2.5pt]{0.02\textwidth}{0.5mm}}), our method ({\color{blue}\rule[2.5pt]{0.02\textwidth}{0.5mm}}) consistently improves the performance of GANs with different parameters.}
		\label{fig:EX_DCGAN}
	\end{figure}

	Furthermore, we use DCGAN\footnote{\label{ft:2} In this part, we do confirmatory experiments using DCGAN with the simple architecture. The dimension of the latent vector is set to 100. Generator and discriminator are all implemented by 4 convolution layers and BN layers. {At last}, we use the Adam optimizer with the most popular hyper-parameters.} to do some comparative experiments on the CIFAR-10 dataset \cite{krizhevsky2009learning}. As shown in Figure \ref{fig:EX_DCGAN}, under different hyper-parameters, the proposed DAT improves the performance of the GANs consistently. Especially when Learning Rate is large (1e-3, 2e-3, and 5e-3), the vanilla training DCGAN cannot be trained due to the gradient vanishing, while DAT-GAN alleviates this situation eminently. The results indicate that our DAT method reduces the sensitivity to hyper-parameters during the training of GANs.
	
	\begin{table*}
		\caption{FID, KID, and IS of several networks across different datasets. IS: higher is better. FID and KID: lower is better. \label{tab:SOTA}}
		\centering
		\scalebox{1}{
			\begin{tabular}{ccccccccc}
				\toprule
				\multirow{2}{*}{Evaluation  set}&\multirow{2}{*}{Metric} &\multirow{2}{*}{Dataset} &\multicolumn{6}{c}{\makecell*[c]{Models}} \\ 
				\cline{4-9} 
				\specialrule{0em}{2pt}{2pt}
				&&&
				\makecell*[c]{SNGAN}&SNGAN-DAT&SSGAN&SSGAN-DAT&InfoMaxGAN&InfoMaxGAN-DAT\\
				\midrule
				\multirow{10}{*}{Train}&\multirow{5}{*}{FID}&{LSUN-Bedroom}&27.26&24.10&23.54&\textbf{20.43}&35.69&34.21\\
				&&{Tiny-ImageNet}&47.14&43.40&40.83&38.98&41.81&\textbf{38.83}\\
				&&{STL-10}&42.62&41.05&39.84&\textbf{35.64}&42.39&40.77\\
				&&{CIFAR-100}&23.70&21.09&22.43&\textbf{19.86}&21.36&20.46\\
				&&{CIFAR-10}&19.75&17.93&16.75&\textbf{15.34}&17.71&17.69\\
				\cline{2-9} 
				\specialrule{0em}{2pt}{2pt}
				&\multirow{5}{*}{KID}&{LSUN-Bedroom}&0.0313&0.0286&0.0267&\textbf{0.0246}&0.0373&0.0369\\
				&&{Tiny-ImageNet}&0.0411&0.0391&0.0355&0.0337&0.0371&\textbf{0.0323}\\
				&&{STL-10}&0.0391&0.0380&0.0388&\textbf{0.0336}&0.0418&0.0384\\
				&&{CIFAR-100}&0.0158&0.0160&0.0155&0.0147&0.0152&\textbf{0.0140}\\
				&&{CIFAR-10}&0.0149&0.0133&0.0125&\textbf{0.0116}&0.0134&0.0128\\
				\cline{1-9} 
				\specialrule{0em}{2pt}{2pt}
				\multirow{10}{*}{Test}&\multirow{5}{*}{FID}&{LSUN-Bedroom}&86.10&83.50&85.10&\textbf{82.20}&87.14&86.17\\
				&&{Tiny-ImageNet}&52.29&48.72&45.79&44.63&47.26&\textbf{43.97}\\
				&&{STL-10}&62.03&60.41&56.56&\textbf{52.71}&61.93&59.07\\
				&&{CIFAR-100}&28.48&26.49&27.61&\textbf{24.67}&26.36&25.28\\
				&&{CIFAR-10}&24.19&22.46&21.23&\textbf{19.45}&22.03&21.88\\
				\cline{2-9} 
				\specialrule{0em}{2pt}{2pt}
				&\multirow{5}{*}{KID}&{LSUN-Bedroom}&0.0324&0.0296&0.0298&\textbf{0.0275}&0.0388&0.0375\\
				&&{Tiny-ImageNet}&0.0415&0.0393&0.0371&0.0341&0.0372&\textbf{0.0332}\\
				&&{STL-10}&0.0445&0.0428&0.0403&\textbf{0.0352}&0.0443&0.0404\\
				&&{CIFAR-100}&0.0164&0.0161&0.0160&0.0148&0.0153&\textbf{0.0142}\\
				&&{CIFAR-10}&0.0150&0.0142&0.0128&\textbf{0.0118}&0.0139&0.0141\\
				
				\cline{1-9}
				\specialrule{0em}{2pt}{2pt}
				\multirow{5}{*}{None}&\multirow{5}{*}{IS}&LSUN-Bedroom&-&-&-&-&-&-\\
				&&Tiny-ImageNet&8.17&8.44&8.63&8.95&8.80&\textbf{9.18}\\
				&&STL-10&8.34&8.45&8.59&\textbf{8.72}&8.28&8.57\\
				&&
				CIFAR-100&7.66&7.82&7.74&\textbf{8.09}&8.02&8.06\\
				&&
				CIFAR-10&7.84&7.97&8.13&\textbf{8.25}&8.01&8.07\\
				\bottomrule
			\end{tabular}
		}
	\end{table*}

	\begin{table*}
		\caption{ Train FID scores on the CIFAR-10 dataset for various GAN losses and regularization methods.\label{tab:fid}}
		\centering
		\begin{tabular}{cccccccccc}
			\toprule
			\multirow{2}{*}{loss} & \multicolumn{9}{c}{\makecell*[c]{regularization}} \\ 
			\cline{2-10} 
			& \makecell*[c]{None}&GP\cite{gulrajani2017improved}&LP\cite{petzka2017regularization}&0-GP&AR\cite{zhou2018don}&RFM\cite{zhou2018don}&ASGAN\cite{liu2020adversarial}&DAT(ours)&DATT(DAT+T) \\ 
			
			\midrule
			GAN\cite{goodfellow2014generative}&  30.20&27.80&27.25&29.76 &29.37&27.89&27.24&26.98&\textbf{26.77} \\
			LSGAN\cite{mao2017least}&25.20  &24.40 &25.38&28.67 &26.49&26.05&26.46&24.38&\textbf{24.35} \\
			WGAN\cite{arjovsky2017wasserstein}&29.43 &26.03&26.90&43.06 &27.26&26.83&25.98&25.67&\textbf{25.66}\\
			\bottomrule
		\end{tabular}
	\end{table*}
	\subsection{Evaluation on some popular methods in different datasets}
	In this section, we apply DAT to some popular methods, such as SNGAN, SSGAN{,} and InfoMAXGAN. The architectures used for all models are equivalent to SNGAN and code can be available on Github\footnote{\url{https://github.com/kwotsin/mimicry}}. We evaluate our method on five different datasets: CIFAR-10, CIFAR-100, STL-10, Tiny-ImageNet, and LSUN-Bedroom. All training parameters are selected with the best results on baseline networks. The details can be found in Table \ref{Tab:parameters}, where the Adam parameters are Learning Rate (LR=2e-4), $\beta_1$=0 and $\beta_2$=0.9; $n_{dis}$ is number of discriminator steps per generator step; $n_{iter}$ is the trained times of the generator. Besides, we use the hinge loss to train all the models. The results are illustrated in Table \ref{tab:SOTA}.
	
	As seen in Table \ref{tab:SOTA}, DAT improves FID, KID, and IS consistently and significantly across many datasets with three popular models, which suggests our method is versatile and can generalize across multiple data domains. For instance, compared to SNGAN, the improvement in Traning FID of SNGAN-DAT is $9.2\%$, $11.0\%$, $3.7\%$, $7.9\%$, $11.6\%$ from CIFAR-10 (32*32), CIFAR-100 (32*32), STL-10 (48*48), Tiny-ImageNet (64*64), to LSUN-Bedroom (64*64). Similarly, compared to SSGAN, the improvement in Traning FID of SSGAN-DAT is $8.4\%$, $11.5\%$, $10.5\%$, $4.5\%$, $13.2\%$ from CIFAR-10, CIFAR-100, STL-10, Tiny-ImageNet, to LSUN-Bedroom.
	
	Due to the large difference between the LSUN-Bedroom distribution and the ImageNet distribution, IS {does} not seem meaningful for the metric of generation on the LSUN-Bedroom dataset. So we do not show IS on the LSUN-Bedroom dataset. 
	
	
	{Furthermore, for qualitative comparisons, we present randomly sampled and non-cherry picked images generated by SSGAN and SSGAN with DAT for CIFAR-10, CIFAR-100, STL-10, and LSUN-Bedroom datasets in section B of the Appendix.}
	
	\begin{figure*}
		\vspace{-3em}
		\includegraphics[width=0.9\textwidth]{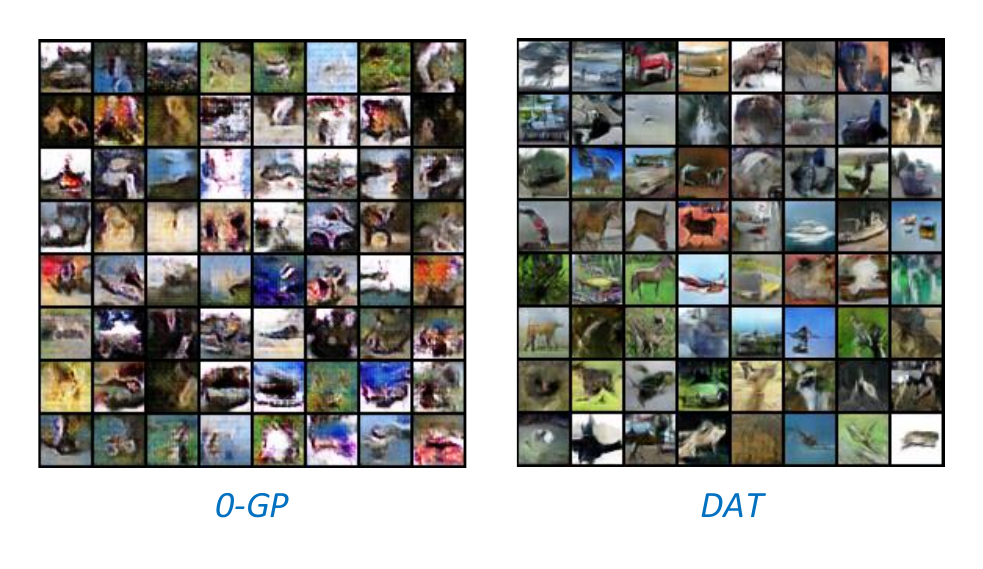}
		\centering
		\vspace{-1em}
		\caption{Randomly generated images with 0-GP (left) and DAT (right) methods under the WGAN loss on the CIFAR-10 dataset. The generated images with 0-GP are unrealistic and contain much artifact, which can be mitigated by the proposed DAT.}
		\label{fig:DAT_0GP}
	\end{figure*}
	
	\subsection{The FID Results on CIFAR-10 Dataset with Other Regularization Methods}
	{From section 4.3, the proposed DAT can be considered as an adaptive gradient minimum (0-GP) method.} To demonstrate the advancements of our method, we compare it with other regularization methods, such as GP, LP, 0-GP, AR, RFM, and ASGAN, on the CIFAR-10 dataset. To make the conclusion more general, we conduct comparative experiments on three loss functions of GANs (GAN \cite{goodfellow2014generative}, LSGAN \cite{mao2017least}, and WGAN \cite{arjovsky2017wasserstein}). The baseline architecture and code adopt the implementation of AR and RFM, which can be available on Github\footnote{\url{https://github.com/bradyz/robust-discriminator-pytorch}}. From which, the dimensions of the latent vector is 128, the batch size is 64, and the Adam optimizer is set to $lr=0.0001$, $\beta_1=0.5$, and $\beta_2=0.999$. 
	
	We have implemented LP, 0-GP, ASGAN, DAT, and DATT based on the baseline code. For instance, LP is a relaxation constraint of GP, and it limits the gradient of the discriminator to $k$ $(k\leq1)$. 0-GP minimizes the gradient of the discriminator, which is similar to the DAT with adaptive gradient minimization. AR and RFM propose the consistent regularization between clean images and adversarial images in the training of GANs, which will let the discriminator not be fooled by adversarial examples. ASGAN is another method using adversarial training for the discriminator. However, like using adversarial training in classification tasks, the adversarial examples in ASGAN are generated by $\hat{x}=x-\epsilon \operatorname{sign}\nabla_{x}(V_m\big(\theta,\phi,x,z)\big)$, which is similar to the untargeted FGSM in Eq \ref{Eq:6}. $V_m\big(\theta,\phi,x,z)$ is the loss function of the discriminator. Here, we use the best-performing hyper-parameter in \cite{liu2020adversarial}, to be more specific, using FGSM on both real and fake samples and the key hyper-parameter $\epsilon=\frac{1}{255}$. DAT represents training the GANs as shown in Algorithm \ref{algorithm:1}, while DATT represents training the network with adversarial examples and normal examples together. Training FID results are illustrated in Table \ref{tab:fid}, which confirm the advancements of our method. Compared to DAT, DATT has {tiny} improvement but requires more training time, which is not cost-effective. We recommend adopting the DAT method to train the GAN model. More specifically, we compare the generation results of 0-GP and DAT under the WGAN Loss, as shown in Figure \ref{fig:DAT_0GP}. We find that the generation results on 0-GP with WGAN loss are {inferior}, which may be caused by the strict limit of minimizing the gradient penalty. DAT will adaptively minimize the gradient, which relieves the above limitation to improve the performance of the generation.
	
	Taking above together, it has been clearly shown that our proposal has more competitive enhancements to those existing efforts for GANs training.
	\begin{figure}
		\includegraphics[width=0.45\textwidth]{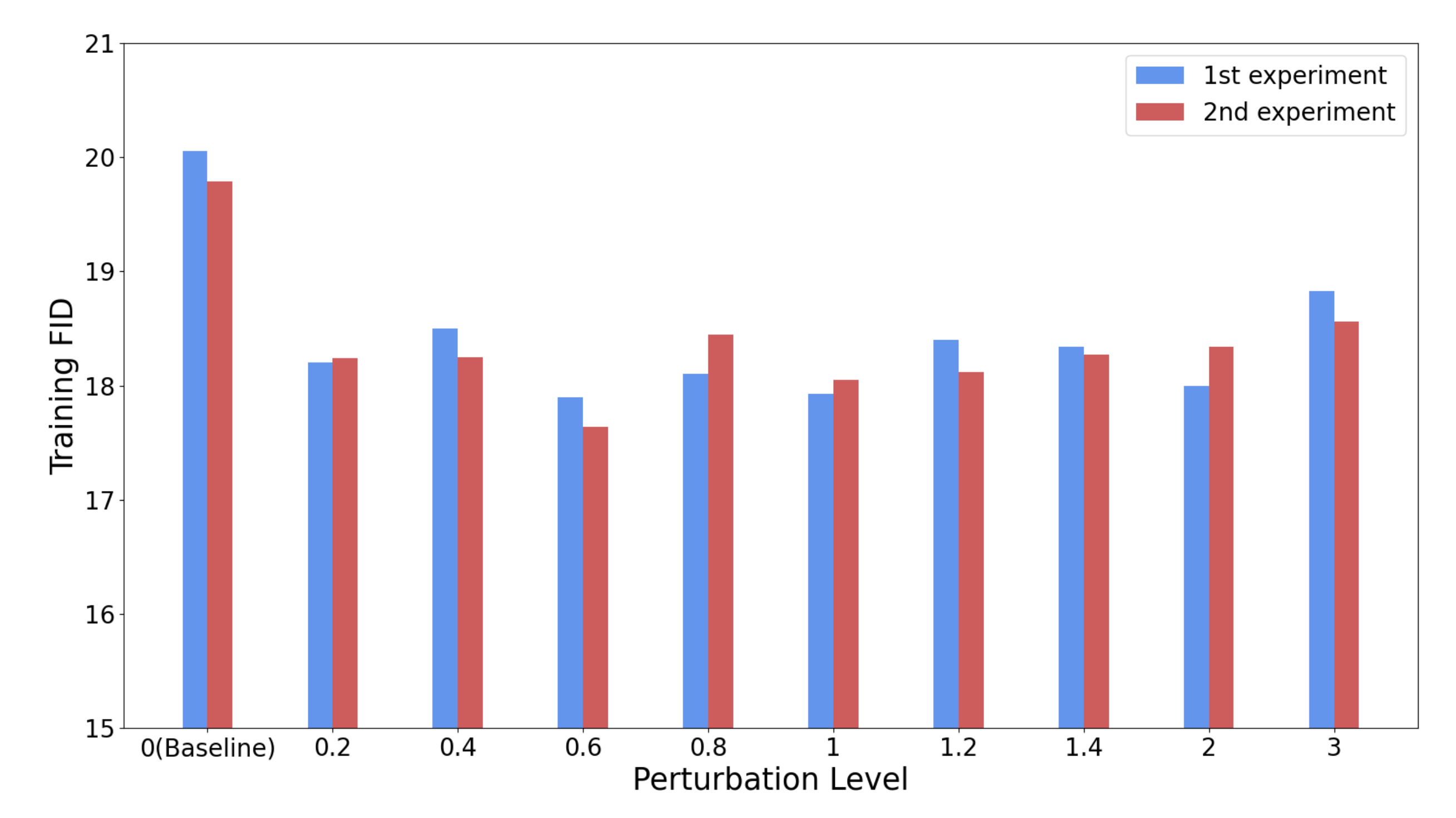}
		\centering
		\caption{Training FID of two independent runs with different perturbation level ($\epsilon$) settings. (Lower is better)}
		\label{fig:perturbation_level}
	\end{figure}
	\begin{figure*}
		\begin{subfigure}[b]{0.45\textwidth}
			\centering
			\includegraphics[width=\textwidth]{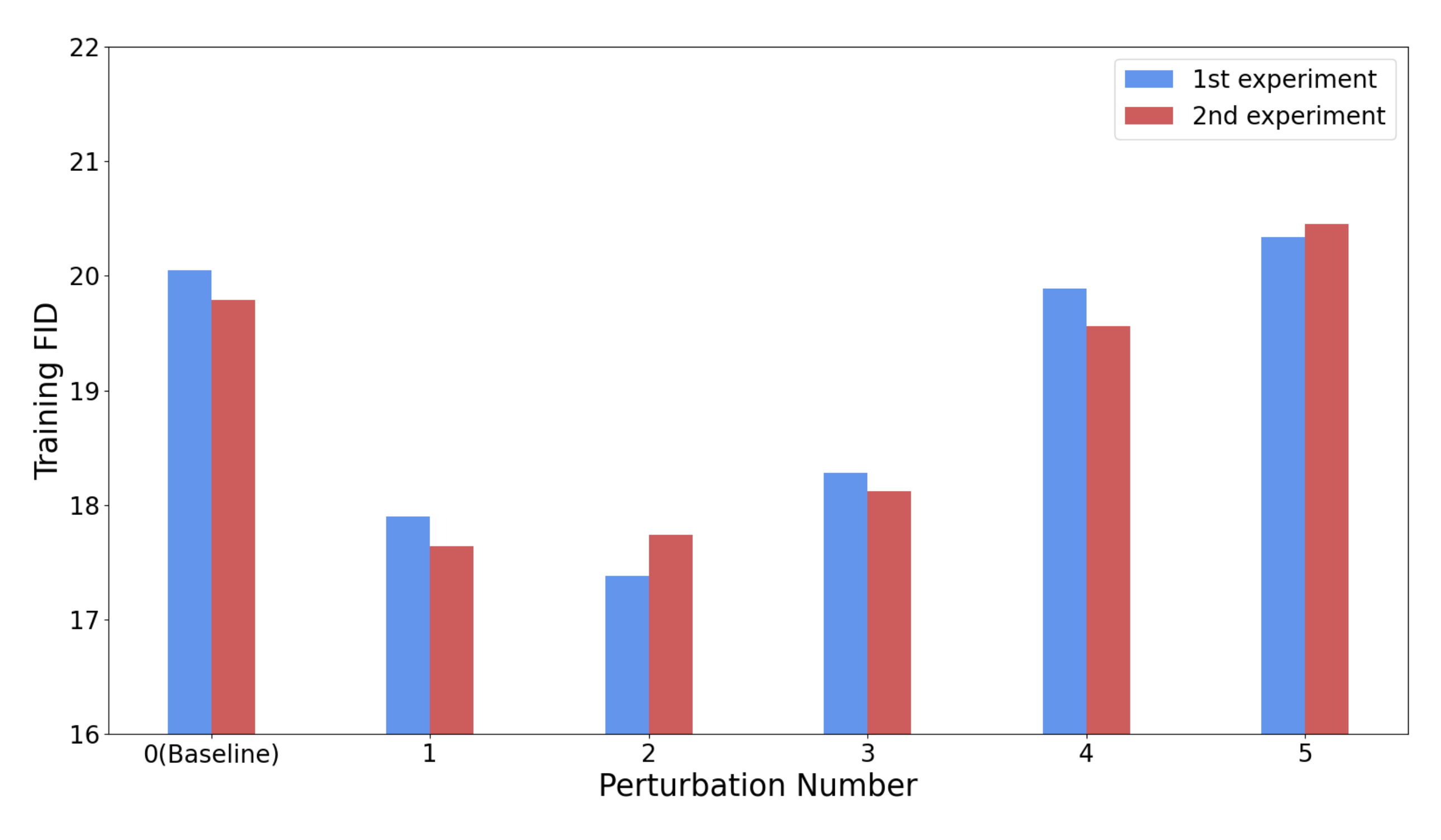}
			\caption{Training FID in different perturbation numbers of two independent runs ($\epsilon=0.6$)}
			
		\end{subfigure}
		\begin{subfigure}[b]{0.45\textwidth}
			\centering
			\includegraphics[width=\textwidth]{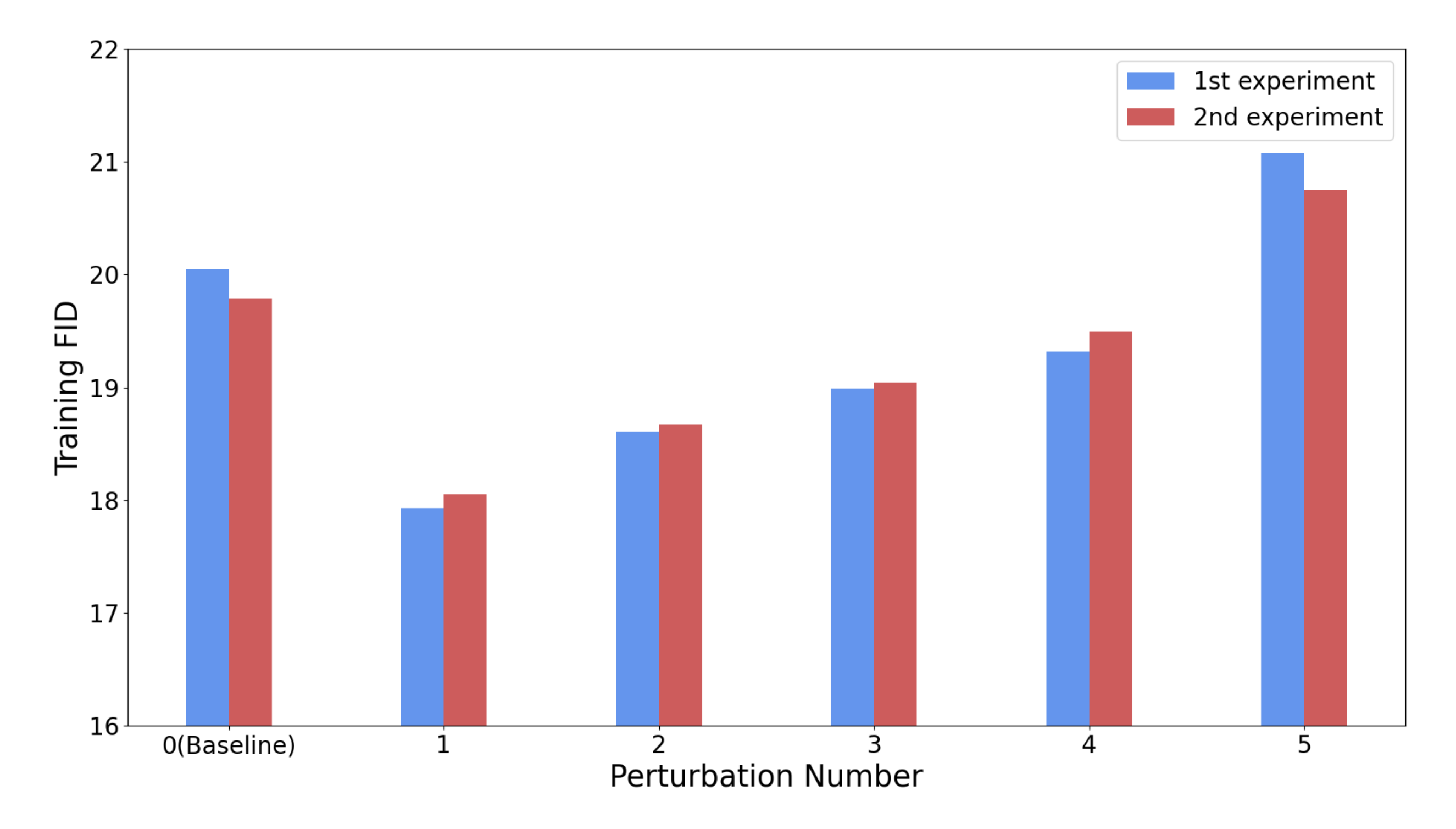}
			\caption{Training FID in different perturbation numbers of two independent runs ($\epsilon=1$)}
			
		\end{subfigure}
		\centering
		\caption{Training FID in different perturbation numbers of two independent runs (Lower is better)}
		\label{figure:perturbation number}
	\end{figure*}
	
	\subsection{The Results on Adversarial Robustness of the Discriminator}
	
\begin{figure}
		\includegraphics[width=0.45\textwidth]{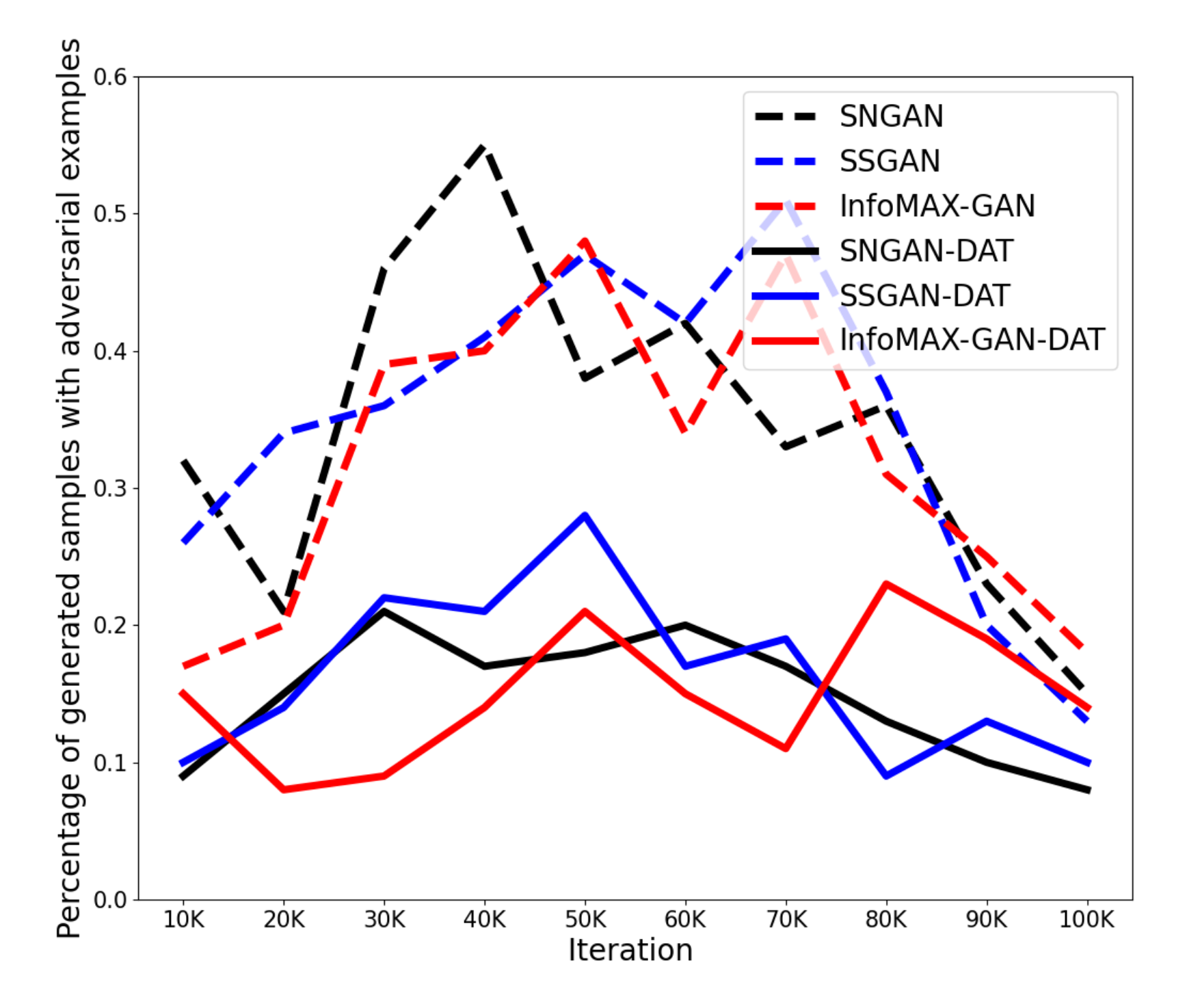}
		\centering
		\caption{Percentage of generated images with adversarial examples on CIFAR-10 dataset. Following the definition of adversarial examples, we set $\delta=0.1$ in this figure. With DAT, the proportion of generated samples existing adversarial samples is low and stable for different GAN models, which stabilizes the training of the GANs. All results are the average of three independent runs. }
		\label{fig:number_adversarial_examples_all}
	\end{figure}

\begin{figure*}
		\begin{subfigure}[b]{0.45\textwidth}
			\centering
			\includegraphics[width=\textwidth]{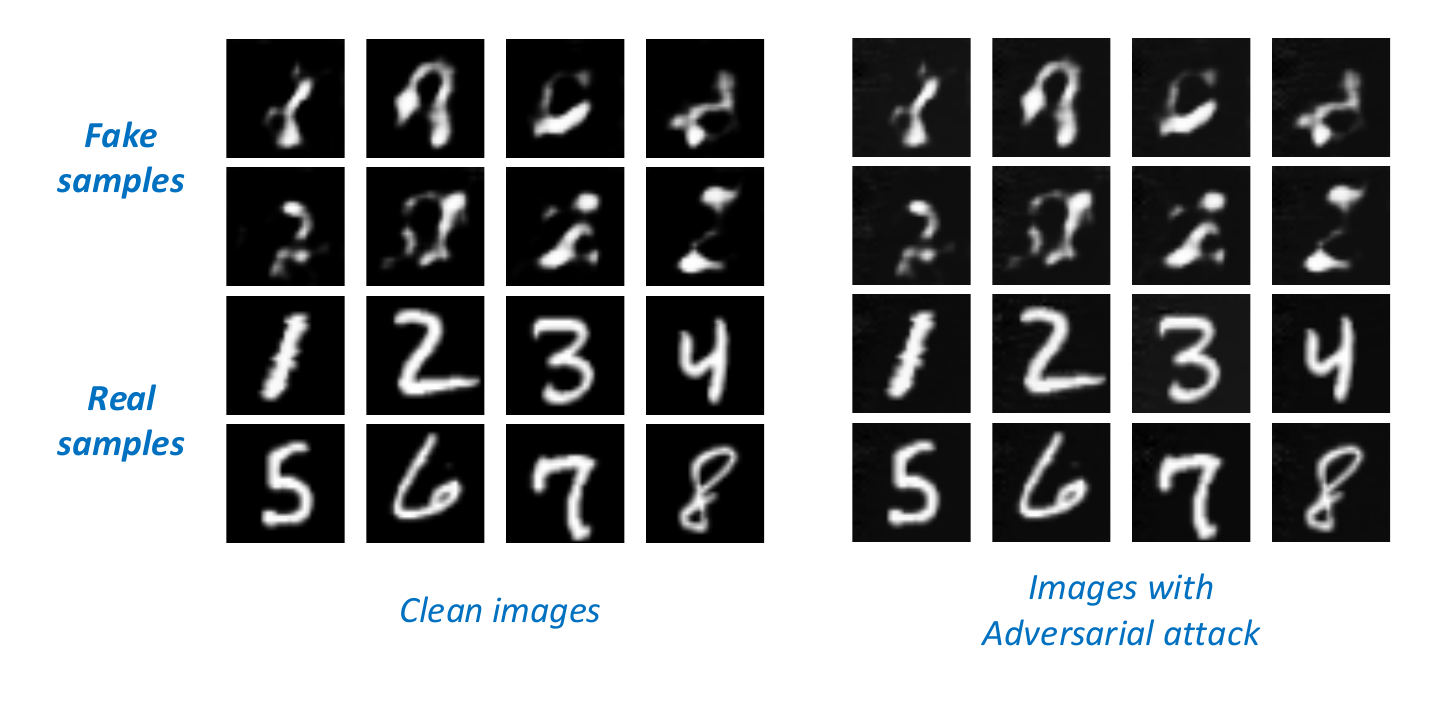}
			\caption{adversarial attack without DAT \protect\\ ($\bar{K}_{fake}=3.2$, $\bar{K}_{real}=3.88$)}
			\label{figure:adv attack without adv training}
		\end{subfigure}
		\begin{subfigure}[b]{0.45\textwidth}
			\centering
			\includegraphics[width=\textwidth]{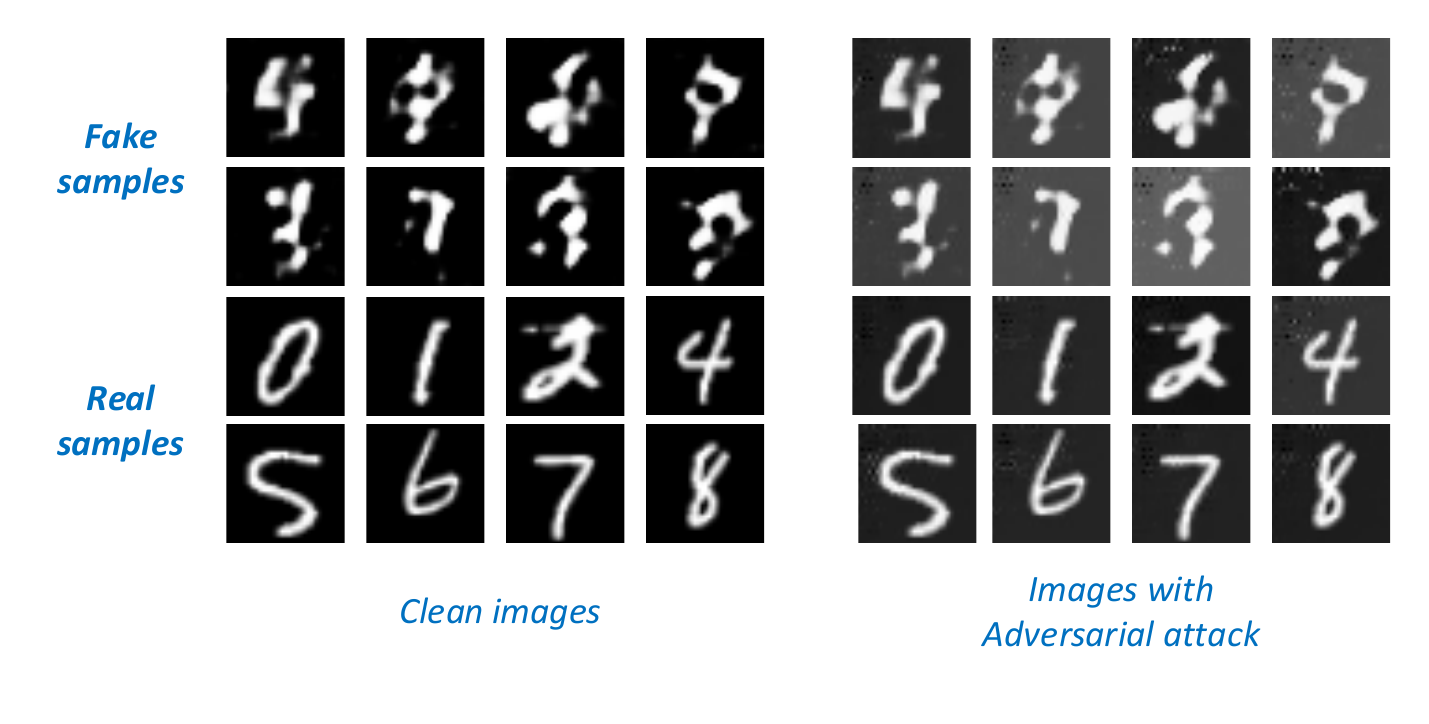}
			\caption{adversarial attack with DAT \protect\\ ($\bar{K}_{fake}=37.36$, $\bar{K}_{real}=7.44$)}
			\label{figure:adv attack with adv training}
		\end{subfigure}
		\centering
		\caption{Adversarial attack of the discriminator. (a) and (b) are the adversarial examples of DCGAN without DAT and with DAT on the MNIST dataset, respectively. Top two lines are the fake samples that are generated by generator without convergence. Bottom two lines are real samples from the MNIST dataset. $\bar{K}_{fake}$ and $\bar{K}_{real}$ are the average number of iterations when fake images and real images successfully attack the discriminator, respectively}
		\label{figure:adv attack}
	\end{figure*}

	{In this section, we illustrate the adversarial robustness of the discriminator. Figure \ref{fig:number_adversarial_examples_all} illustrates the proportion of generated samples existing adversarial samples with and without DAT on the CIFAR-10 dataset. DAT can significantly reduce this proportion. Furthermore, we also show the robust to adversarial attack of the trained discriminator in Figure \ref{figure:adv attack}. Figure \ref{figure:adv attack} (a) illustrates that average iterations for successfully attacking a discriminator without DAT through PGD are 3.2 and 3.88, which indicates that vanilla GAN models can be easily attacked. Also, we cannot distinguish between clean images and adversarial examples intuitively. Average iterations of the successful attack are significantly reduced when DAT is used in Figure \ref{figure:adv attack} (b). }
	
	\section{Exploratory and Ablation Studies}
	\subsection{Evaluation with different perturbation levels}
	Similar to other adversarial training, perturbation level ($\epsilon$ in Eq (\ref{Eq:17})) is also important for DAT. In this section, we evaluate the performance of the DAT with different $\epsilon$ settings on the CIFAR-10 dataset. Specifically, we perform unconditional image generation with SNGAN architecture in different settings of perturbation level. We select certain classic perturbation levels for experiments such as \{0.2, 0.4, 0.6, 0.8, 1, 1.2, 1.4, 2, 3\}. {All experiments are conducted twice independently to reduce the effect of randomness.}
	
	As shown in Figure \ref{fig:perturbation_level}, the performance of the DAT is robust for different settings of $\epsilon$, and the optimal setting of $\epsilon$ is around 0.6. We analyze the impact of $\epsilon$ from two perspectives. From the perspective of adversarial training, the improvement of the original model by a small level of perturbation is negligible. In this case, the impact of adversarial training is finite. On the contrary, strong perturbation levels affect the distribution of the images and degrade the performance compared to the optimal parameter settings. Furthermore, from the perspective of gradient penalty, perturbation level is the coefficient of the gradient penalty, as shown in Eq \ref{Eq:12} and Eq \ref{Eq:15}. A tiny penalty relaxes the restriction on the discriminator's Lipschitz continuity, while a strong penalty leads to the tight restriction of the discriminator's Lipschitz continuity. Both of them affect the performance of GANs. 
	\begin{figure*}
		\includegraphics[width=0.9\textwidth]{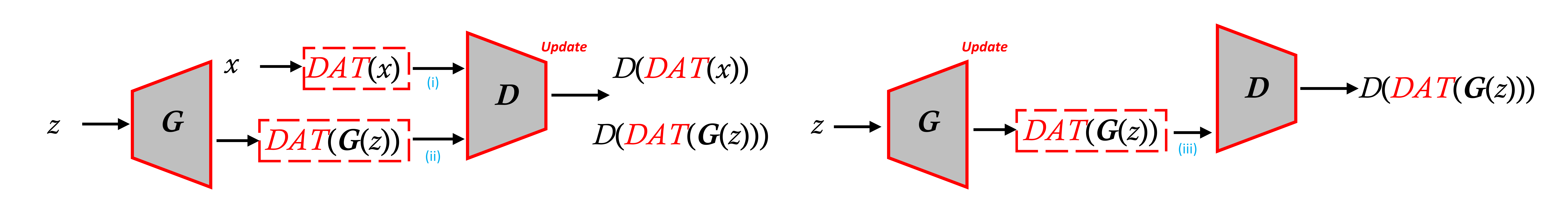}
		\centering
		\caption{The updating of the discriminator (left) and generator(right) with DAT. }
		\label{fig:overview_of_the_dat}
	\end{figure*}
	\begin{table*}
		\caption{The results using DAT in different perturbation position for training the GANs on CIFAR-10 dataset. "Generator only" uses DAT to position (iii) (see Figure \ref{fig:overview_of_the_dat});
			"Real only" uses DAT to position (i); "Fake only" uses DAT to position (ii); "Discriminator only"  uses DAT to both position (i) and position (ii), but not position (iii); "All" uses DAT to position (i), (ii), and (iii). "Discriminator only" is described in section 4 and is applied in other sections. We select the best FID results for each method during the training.
			\label{Tab:position}}
		\centering
		\begin{tabular}{cccccccccc}
			\toprule
			\multirow{3}{*}{Methods}&\multicolumn{3}{c}{Where { $DAT$}? }&\multicolumn{5}{c}{Results}\\ 
			\cline{2-4} 	\cline{6-10}
			&
			\makecell*[c]{(i)}&(ii)&(iii)
			&&IS&Train\_FID&Test\_FID&Train\_KID&Test\_KID\\
			
			\midrule
			SNGAN(Baseline)&&&&&7.84&19.75&24.19&0.0149&0.0150\\
			\midrule
			SNGAN-DAT (Generator only)&&&\checkmark&&1.59&309&310.5&0.3836&0.3822\\
			SNGAN-DAT (Real only)&\checkmark&&&&7.86&19.48&23.73&0.0147&0.0149\\
			SNGAN-DAT (Fake only)&&\checkmark&&&7.84&19.64&24.1&0.015&0.0155\\
			{SNGAN-DAT (Discriminator only)}&\checkmark&\checkmark&&&{7.97}&{17.93}&{22.46}&{0.0133}&{0.0142}\\
			SNGAN-DAT (All)&\checkmark&\checkmark&\checkmark&&1.62&292&294&0.3603&0.3605\\
			\bottomrule
		\end{tabular}
	\end{table*}
	
	\subsection{Evaluation with different perturbation numbers}
	As acknowledged, in addition to the one-step adversarial perturbation mentioned in Section 4, adversarial training can be implemented by multi-step methods, such as PGD. Generally, these multi-step methods can also be used to generate adversarial examples in DAT. Certainly, different perturbation numbers are expected to get more attractive performance for GANs. In this section, we experiment the DAT method with different number of adversarial perturbations on the CIFAR-10 dataset, where the basic architecture is SNGAN and the perturbation level ($\epsilon$) are set to 0.6 or 1. We select certain classic perturbation numbers for experiments such as \{1, 2, 3, 4, 5\}. All experiments are performed two times to reduce the effect of randomness.
	
	As shown in Figure \ref{figure:perturbation number}, the left part illustrates the Train FID with $\epsilon=0.6$ and the right part illustrates the Train FID with $\epsilon=1$. DAT with multiple perturbation numbers achieves more significant improvement when the perturbation level is set to 0.6. However, the performance of GANs decreases with the number of perturbations when the perturbation level is set to 1. The results also demonstrate that neither too large nor too small adversarial perturbations are effective in improving the performance of GANs. For instance, the best performance has been achieved when perturbation level is set to 0.6 and perturbation number is set to 2. 
	
	\subsection{Evaluation with different perturbation positions}
	Different from the perturbation in classifiers, adversarial perturbation in GANs can be applied to different positions, which have different effects. In this section, we experiment the DAT with different positions of adversarial perturbation on the CIFAR-10 dataset, where the base architecture is SNGAN, the perturbation level, and the perturbation number are set to 1. Figure \ref{fig:overview_of_the_dat} illustrates the updating of discriminator and generator with the DAT during GANs training, where (i), (ii), (iii) are different perturbation positions for DAT. The experimental results in Table \ref{Tab:position} illustrate that applying DAT to the generator's training ("Generator only" and "All" methods) destroys the training of GANs, which is noticeable. Generator is trained by information from discriminator. During generator training, adversarial training is not useful and breaks the gradient. Furthermore, using DAT only for fake images ("Fake only" method) or real images ("Real only" method) does not significantly improve the performance of GANs training. {As described in Algorithm 1, the best performance is using DAT for real and fake images during discriminator training.}
	
	\section{Conclusions}
	Motivated by adversarial examples in neural networks, in this paper, we argue that the instability of GANs training has close relation with adversarial examples of the discriminator. {Unlike} the adversarial perturbation in classifiers, adversarial noise of the discriminator produces the incorrect gradient of the generator, which misleads the update of the generator and causes the instability of GANs training. With this discovery, we propose the DAT method to avoid the adversarial examples during discriminator training. {Furthermore, the proposed DAT method can adaptively adjust the Lipschitz constants, superior to some fixed gradient penalty methods, such as GP, LP, and 0-GP.} The validation with different architectures on varied datasets indicates that the proposed DAT method can significantly improve the performance and alleviate mode collapse. In addition, some ablation studies compare the effect of different perturbation levels, perturbation numbers, and perturbation positions. {The results indicate that more complex adversarial manipulation methods can further improve the DAT's performance.}
	
	\section*{Acknowledgment}
	
	The work is partially supported by the National Natural Science Foundation of China under grand No.U19B2044, No.61836011.


	\ifCLASSOPTIONcaptionsoff
	\newpage
	\fi
	
	\bibliographystyle{IEEEtran}
	\bibliography{bare_jrnl_transmag}
	\begin{IEEEbiography}[{\includegraphics[width=1in,height=1.25in,clip,keepaspectratio]{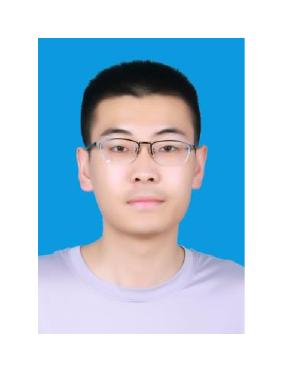}}]{Ziqiang Li}
		received the B.E. degree from University
		of Science and Technology of China (USTC), Hefei,
		China, in 2019 and is pursing the Ph.D. degree
		from University of Science and Technology of China
		(USTC), Hefei, China. His research interests include
		medical image segmentation,deep generative models, and machine learning.
	\end{IEEEbiography}
	\begin{IEEEbiography}[{\includegraphics[width=1in,height=1.25in,clip,keepaspectratio]{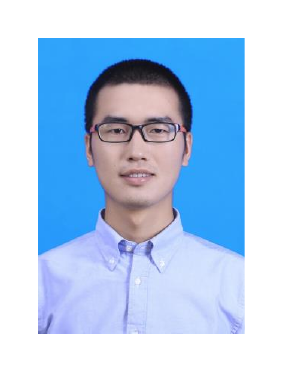}}]{Rentuo Tao}
		received the B.E. degree from Hefei
		University of Technology (HFUT), Hefei, China, in
		2013 and is pursing the Ph.D. degree from University of Science and Technology of China (USTC),
		Hefei, China. His research interests include deep
		generative models, machine learning and computer
		vision.
	\end{IEEEbiography}
	\begin{IEEEbiography}[{\includegraphics[width=1in,height=1.25in,clip,keepaspectratio]{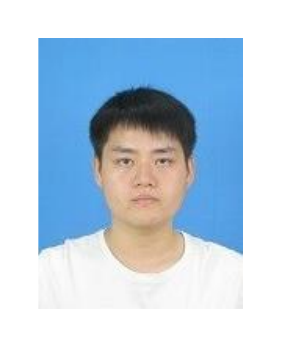}}]{Pengfei Xia}
		received the B.E. degree from the China University of Mining and Technology (CUMT), XuZhou, China, in 2015. He is currently pursuing the Ph.D. degree with the University of Science and Technology of China (USTC), Hefei, China. His research interests including adversarial examples and deep learning security.
	\end{IEEEbiography}
	\begin{IEEEbiography}[{\includegraphics[width=1in,height=1.25in,clip,keepaspectratio]{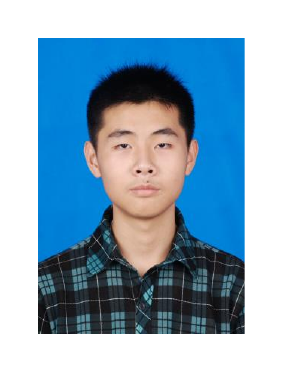}}]{Hongjing Niu}
		received the B.E. degree from University
		of Science and Technology of China (USTC), Hefei,
		China, in 2018 and is pursing the PHD degree
		from University of Science and Technology of China
		(USTC), Hefei, China. His research interests include
		interpretation of deep network,visualization and machine learning.
	\end{IEEEbiography}
	\begin{IEEEbiography}[{\includegraphics[width=1in,height=1.25in,clip,keepaspectratio]{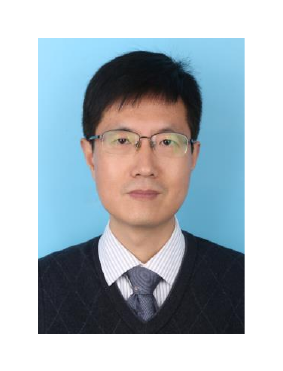}}]{Bin Li}
		received the B.Sc. degree from the Hefei
		University of Technology, Hefei, China, in 1992, the
		M.Sc. degree from the Institute of Plasma Physics,
		Chinese Academy of Sciences, Hefei, in 1995, and
		the Ph.D. degree from the University of Science and
		Technology of China (USTC), Hefei, in 2001. He is
		currently a Professor with the School of Information
		Science and Technology, USTC. He has authored
		or co-authored over 40 refereed publications. His
		current research interests include evolutionary computation, pattern recognition, and humancomputer
		interaction. Dr. Li is the Founding Chair of the IEEE Computational Intelligence Society Hefei Chapter, a Counselor of the IEEE USTC Student Branch,
		a Senior Member of the Chinese Institute of Electronics (CIE), and a member
		of the Technical Committee of the Electronic Circuits and Systems Section
		of CIE.
	\end{IEEEbiography}
	
	\clearpage
	\appendix

	\section{Appendix}
	
	\newcommand{\topcaption}{%
		\setlength{\abovecaptionskip}{0pt}%
		\setlength{\belowcaptionskip}{0pt}%
		\caption}
	\subsection{Evaluation Settings}
	In this paper, we use three metrics: Fréchet Inception Distance (FID) \cite{heusel2017gans}, Kernel Inception Distance (KID) \cite{binkowski2018demystifying}, and Inception Score (IS) \cite{salimans2016improved} to evaluate the quality and diversity of generated images.
	\subsubsection{Fréchet Inception Distance} 
	FID is the most popular metric evaluating the performance of GANs, which is widely used in the literature. FID calculates the Wasserstein-2 distance of features produced by pre-trained Inception network between real and generated images. Formally, FID is defined as:
	\begin{equation}
		d_{\mathrm{FID}}=\left\|\mu_{r}-\mu_{g}\right\|_{2}^{2}+\operatorname{Tr}\left(\Sigma_{r}+\Sigma_{g}-2\left(\Sigma_{r} \Sigma_{g}\right)^{\frac{1}{2}}\right),
	\end{equation}
	where $\mu_r$ and $\mu_g$ denote the mean of feature vectors produced by real and fake images, respectively, and $\Sigma_{r}$ and $\Sigma_{g}$ denote the covariance of feature vectors produced by real and fake images, respectively. However, FID can produce biases, where the FID between train and generated images have better scores, and FID between test and generated images have worse scores.  
	\subsubsection{Kernel Inception Distance}
	KID is also a alternative metric of FID, which produces small bias. Formally, KID measures the square of the Maximum Mean Discrepancy (MMD) between two distributions:
	\begin{equation}
		\begin{aligned}
			d_{\mathrm{KID}} &=\operatorname{MMD}^{2}(X, Y) \\
			&=\frac{1}{m(m-1)} \sum_{i \neq j}^{m} k\left(x_{i}, x_{j}\right) \\
			&+\frac{1}{n(n-1)} \sum_{i \neq j}^{n} k\left(y_{i}, y_{j}\right)-\frac{2}{m n} \sum_{i=1}^{m} \sum_{j=1}^{n} k\left(x_{i}, y_{j}\right), \\
		\end{aligned}
	\end{equation}
	where $X$ and $Y$ represent real and generated distribution, respectively, $m$ and $n$ are sample sizes from real and generated distribution, and $k$ is the polynomial kernel defined by:
	\begin{equation}
		k(x, y)=\left(1+\frac{1}{d} x^{T} y\right)^{3}.
	\end{equation}
	Furthermore, the feature of real and generated images is also produced by pre-trained Inception network.
	\subsubsection{Inception Score}
	Different from above methods measuring the distance between real and generated distributions, IS only uses generated images to produce scores. IS measures the performance of GANs in terms of both quality and diversity of images. Quality of generated images is represented by the entropy of the conditional class distribution: $H(p(y|x))$. The smaller $H(p(y|x))$ indicates the higher certainty of the current image classification, which demonstrates the better quality of the generated images. Furthermore, diversity of images is represented by the entropy of marginal class distribution: $H(p(y))$. The larger $H(p(y))$ indicates the generation of category-balanced images, which demonstrates the better diversity of the generated images. In summary, IS can be formally defined as:
	\begin{equation}
		d_{I S}=\exp \left(\mathbb{E}_{x \sim p_{g}} \mathcal{D}_{\mathrm{KL}}(p(y \mid x) \| p(y))\right).
	\end{equation}
	Also, Both the $H(p(y|x))$ and $H(p(y))$ are computed by a pre-trained Inception network.
	\subsection{ Generated Image Samples}
\begin{figure*}
		\begin{subfigure}[b]{0.8\textwidth}
			\centering
			\includegraphics[width=\textwidth]{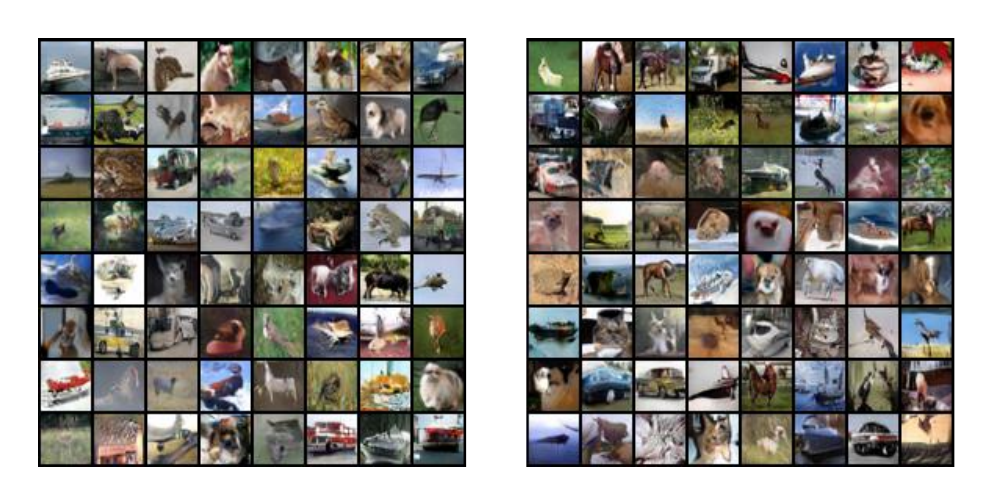}
			\topcaption{CIFAR10}
			\label{figure:adv attack without adv training}
		\end{subfigure}
		\begin{subfigure}[b]{0.8\textwidth}
			\centering
			\includegraphics[width=\textwidth]{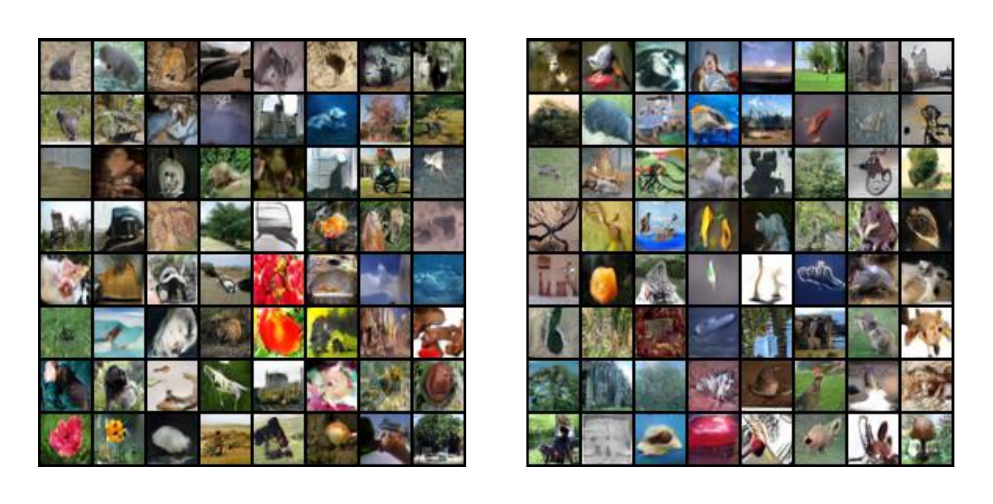}
			\topcaption{CIFAR100}
			\label{figure:adv attack with adv training}
		\end{subfigure}
		\begin{subfigure}[b]{1\textwidth}
			\centering
			\includegraphics[width=0.8\textwidth]{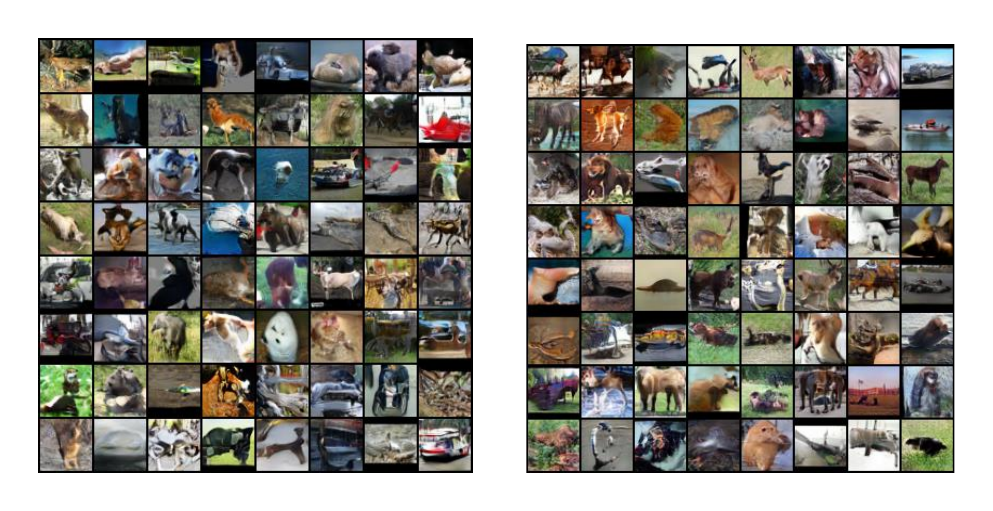}
			\topcaption{STL10}
			\label{figure:adv attack with adv training}
		\end{subfigure}
		\centering
		\caption{Randomly sampled images generated by SSGAN (left) and SSGAN-DAT (right) on CIFAR-10, CIFAR-100, and STL-10 datasets.
		}
		\label{figure:sampled_images_1}
	\end{figure*}

\begin{figure*}
		\begin{subfigure}[b]{0.62\textwidth}
			\centering
			\includegraphics[width=\textwidth]{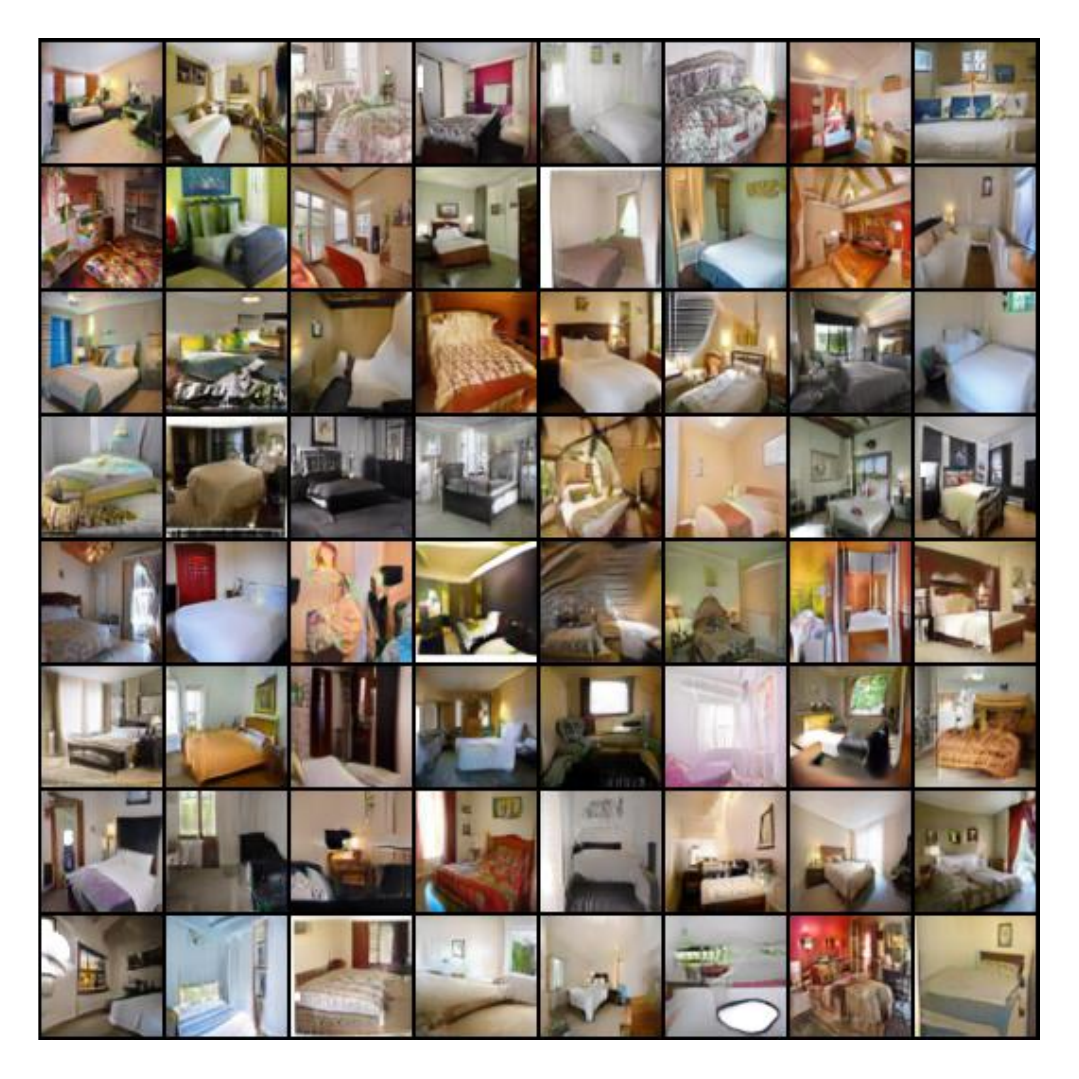}
			\topcaption{SSGAN}
			\label{figure:adv attack without adv training}
		\end{subfigure}
		\begin{subfigure}[b]{0.62\textwidth}
			\centering
			\includegraphics[width=\textwidth]{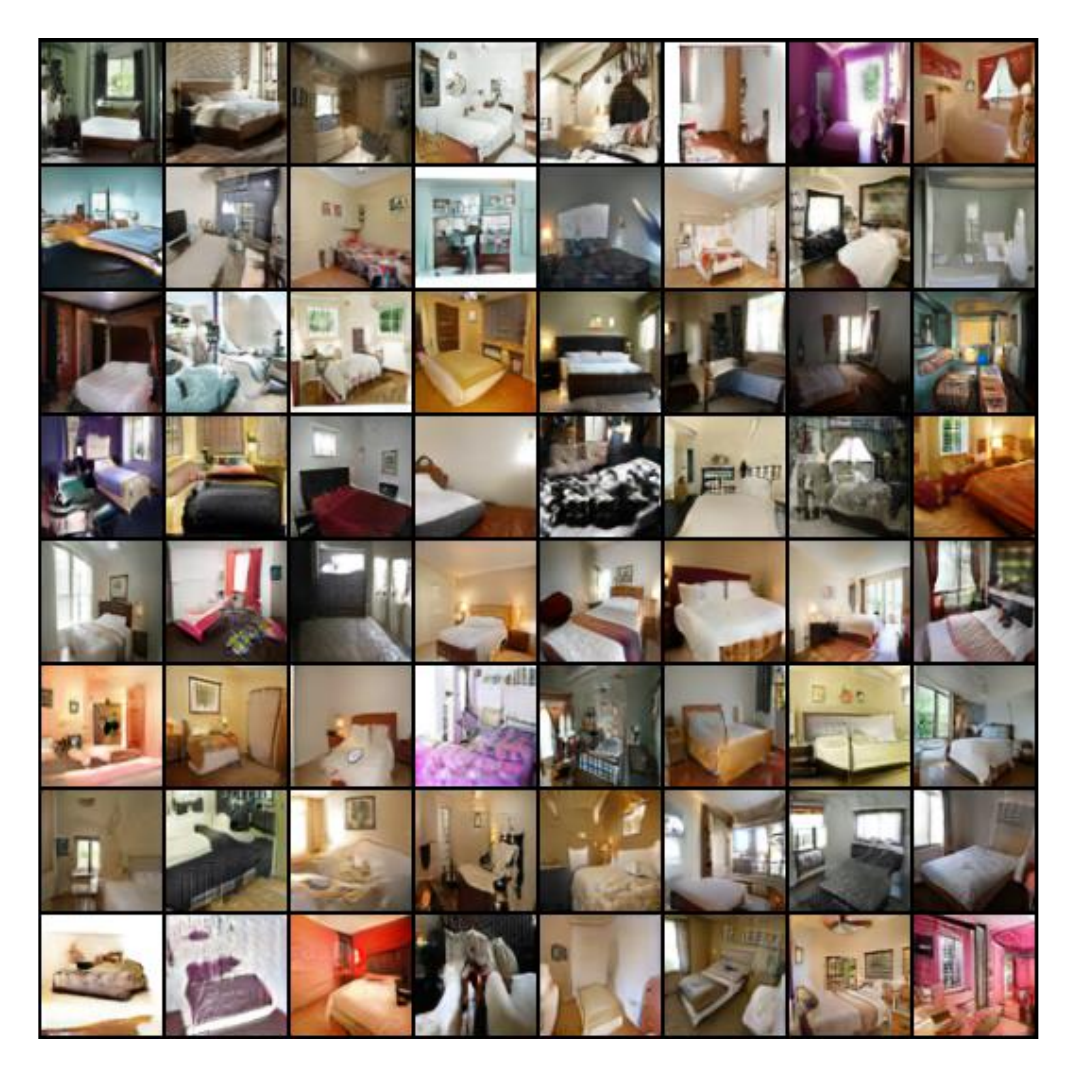}
			\topcaption{SSGAN-DAT}
			\label{figure:adv attack with adv training}
		\end{subfigure}
		\centering
		\caption{Randomly sampled images generated by SSGAN (top) and SSGAN-DAT (bottom) on LSUN-Bedroom dataset.
		}
		\label{figure:sampled_images_2}
	\end{figure*}

	In this section, we show some sampled images generated by SSGAN and SSGAN-DAT on CIFAR-10, CIFAR-100, STL-10, and LSUN-Bedroom dataset in Figures \ref{figure:sampled_images_1} and \ref{figure:sampled_images_2}. The results demonstrate that the images generated images are more diverse and have higher quality after the use of DAT.
	
\end{document}